\documentclass[pra,twocolumn,floatfix]{revtex4-1}

\usepackage{latexsym} 
\usepackage{amssymb}  
\usepackage{amsfonts} 
\usepackage{amsbsy}   
\usepackage{amsmath} 
\usepackage{graphicx}       
\usepackage{subfigure}  
\usepackage{multirow}  
\usepackage{color} 
\usepackage{bm}   
\usepackage{CJK}

\newcommand{\p}{\partial}
\newcommand{\<}{\langle}
\renewcommand{\>}{\rangle} 
\newcommand{\txt}{\textstyle}
\newcommand{\dsp}{\displaystyle}
\newcommand\Eqn[1]{Eq.~(\ref{#1})}  
\newcommand{\beq}{\begin{equation}}
\newcommand{\eeq}{\end{equation}}
\newcommand{\bea}{\begin{eqnarray}}
\newcommand{\eea}{\end{eqnarray}}
\newcommand{\ba}{\begin{array}}
\newcommand{\ea}{\end{array}}

\newcommand{\half} {{\txt \frac{1}{2}}}

\newcommand{\e}{{\rm e}}   
\newcommand{\ri}{{\rm i}}
\newcommand{\rd}{{\rm d}}

\newcommand{\nn}{\nonumber}

\renewcommand{\Re}{{\rm Re}\,}

\newcommand{\ie}{{i.e.}}

\makeatletter 

\begin{document}
	
\title{\bf Piecewise Adiabatic Following in Non-Hermitian Cycling}

\author{Jiangbin Gong}
\affiliation{Department of Physics, National University of Singapore, 117542, Singapore}

\author{Qing-hai Wang}
\affiliation{Department of Physics, National University of Singapore, 117542, Singapore}
	
\date{May 18, 2018}

\begin{abstract}
The time evolution of periodically driven non-Hermitian systems is in general non-unitary but can be stable.  It is hence of considerable interest to examine the adiabatic following dynamics  in periodically driven non-Hermitian systems. We  show in this work the possibility of \textit{piecewise} adiabatic following interrupted by hopping between instantaneous system eigenstates.   This phenomenon is first observed in a computational model and then theoretically explained, using an exactly solvable model, in terms of the Stokes phenomenon.  In the latter case, the piecewise adiabatic following is shown to be a genuine critical behavior and the precise phase boundary in the parameter space is located.  Interestingly, the critical boundary for piecewise adiabatic following is found to be unrelated to the domain for exceptional points. To characterize the adiabatic following dynamics, we also advocate a simple definition of the Aharonov-Anandan (AA) phase for non-unitary cyclic dynamics, which always yields real AA phases.  In the slow driving limit, the AA phase reduces to the Berry phase if adiabatic following persists throughout the driving without hopping, but oscillates violently and does not approach any limit in cases of piecewise adiabatic following.  This work exposes the rich features of non-unitary dynamics in cases of slow cycling and should stimulate future applications of non-unitary dynamics.
\end{abstract}

\begin{titlepage}
	\maketitle
	\renewcommand{\thepage}{}          
\end{titlepage}

\section{Introduction}
\label{sec:intro}

Spectral and dynamical aspects of non-Hermitian systems have attracted considerable theoretical and experimental interests \cite{BB98,PTreview07,GongWang10,PTtheory,PTexp}. Such systems may be regarded as certain extensions of quantum mechanics or phenomenological descriptions of open quantum systems \cite{newRef1}, but more often they model realistic systems in the classical domain with loss and gain, such as waveguides \cite{s0,s1}, LRC circuits \cite{LRC}, mechanical oscillators \cite{s2,s3,s4,s5,s6}, as well as acoustic systems \cite{acoustic1,acoustic2}.  However, little is known about the dynamics of these systems when it is non-unitary. A notable exception was the recent discovery about the impossibility to achieve the adiabatic following due to the circling around exceptional points \cite{BU,Uzdin,Rotter}.

The non-unitary dynamics of non-Hermitian systems is thus expected to be highly useful to further explore and extend the physics underlying the quantum adiabatic theorem \cite{at}.  Indeed, we shall expose in this work a new face of adiabatic following dynamics, featured by a remarkable hopping behavior in system's eigen-representation. This hopping yields piecewise adiabatic following.  Whether or not such piecewise adiabatic following occurs is determined by a phase boundary in the parameter space, thus identified as a true critical behavior.

Below we work on periodically driven non-Hermitian systems, where non-unitary but stable time evolution is recently shown to be possible \cite{Joglekar,GongWang15}.  Thanks to this stable nature of a wide class of non-Hermitian systems, it is convenient to adopt conventional quantum mechanics concepts and tools to explore periodically driven non-Hermitian systems.  In particular, recognizing the importance of geometrical phases in physics in general \cite{book},  we shall examine closely the behavior of the Aharonov-Anandan (AA) phase \cite{AA87} in non-unitary cyclic dynamics to characterize the geometrical aspects of adiabatic following.  It is well known that the Berry phase reflects the geometry of instantaneous Hamiltonian eigenstates in a projective Hilbert space \cite{Berry84}, whereas the AA phase \cite{AA87} reflects the geometry of a curve in a projective Hilbert space traced out by actual cyclic time evolution (one important application of the AA phase is nonadiabatic holonomic quantum computation \cite{zhu}).  By definition then,  the Berry phase and the AA phase in the slow driving limit can be regarded as the same in the presence of ideal adiabatic following.

For a Hermitian Hamiltonian varies sufficiently slowly, then the system may always remain in an instantaneous energy eigenstate. In this case, the AA phase reduces to the Berry phase in the slow driving limit. However, anticipating the possible breakdown of adiabatic following in non-Hermitian systems, one should not pre-assume any correspondence between the Berry phase and the AA phase in the slow driving limit.  To address this issue and also to use the AA phase to depict non-unitary dynamics, we surveyed the literature devoted to generalizing the geometry phases to cases with non-unitary cycling. For example, Samuel and Bhandari generalized Berry phase in the most general setting and obtained always real Berry phases \cite{Samuel88}. Mostafazadeh proposed a real Berry phase using the duel eigenstates in the biorthonormal basis \cite{Ali99}.  Complex Berry phases or complex AA phases in non-Hermitian systems were obtained or positively discussed in other studies \cite{GW88,Aichison92,Wu96,Child97,Child98,Zheng15,Maamache15}.  As elaborated in Sec.~\ref{sec:revisit}, consistent with the work by Samuel and Bhandari \cite{Samuel88},  we adopt  a definition for the AA phase that can be regarded as a natural extension of a previous AA phase expression \cite{Child97}.  Our definition for the AA phase always yields real phases, which can be regarded as one side contribution of this work and lays an excellent foundation for our quantitative investigations below.

Three specific periodically driven non-Hermitian systems are studied in this work. In the first exactly solvable model, it is seen that as the driving period increases, the AA phase smoothly approaches the Berry phase. This feature is a clear indicator that there is no issue to assume adiabatic following.  In the second model, in one parameter regime it behaves as in the first model, but in other parameter regimes,  the AA phase is found to violently oscillate without having a limit at all in the slow driving limit. This is explained by  computationally observing an exotic hopping phenomenon in representation of instantaneous system eigenstates.   To develop more theoretical understandings,  we turn to the third model, the exactly solvable Berry-Uzdin model \cite{BU}, but investigated differently than before.  Using the third model, we are able to show theoretically that the piecewise adiabatic following interrupted by a hopping behavior originates from a phase transition in the parameter space.  We explain the phase transition boundary by the Stokes phenomenon.  Somewhat surprisingly,  the breakdown of adiabatic following in our models occurs in a regime far away from the exceptional points in the parameter space, thereby distinguishing our findings from previous discussions.

This paper is organized as follows. In Sec.~\ref{sec:revisit}, we provide an original framework regarding an always real definition of the AA phase. In Sec.~\ref{sec:solvable}, we present our analysis for our first exactly solvable model.  We then computationally study a second model in Sec.~\ref{sec:hopping}.  To develop a theory of piecewise adiabatic following  based on our observations made in Sec.~\ref{sec:hopping}, we theoretically study the Berry-Uzdin model in Sec.~\ref{sec:BU}. Sec.~\ref{sec:conclusions} concludes this work. 

\section{Revisiting AA phase in non-unitary dynamics}
\label{sec:revisit}
\subsection{General considerations}

Consider a general time-evolving state $|\psi(t)\>$ being cyclic at $t=T$, with $|\psi(T)\>=e^{\ri\alpha}|\psi(0)\>$. The context of $|\psi(t)\>$ and its detailed time dependence (may satisfy a linear or even nonlinear equation of motion) are not needed here.  Because the dynamics under consideration is non-unitary in general, $\alpha$ can be complex. The above-defined phase factor $\alpha$ can be expressed as
\begin{equation}
\alpha = \frac{1}{\ri} \ln \frac{\<\psi(0)|\psi(T)\>}{\<\psi(0)|\psi(0)\>}.
\end{equation}
In this paper, we adopt a single base notation with the bra states being defined as the Hermit conjugate of the ket states, $\<\cdot|\equiv|\cdot\>^\dag$. This is simpler than the use of bi-orthonormal basis adopted in some literature. 
The normalization of $|\psi(t)\>$ is of no interest in developing geometrical insights into the dynamics. Indeed, since the geometry in the projective Hilbert space is of the main concern, there is no reason to be particularly interested in the time dependence of the normalization of $|\psi(t)\>$. Associated with $|\psi(t)\>$, we now define a normalized time-evolving state as
\begin{equation}
|\phi(t)\>\equiv \frac{|\psi(t)\>}{\sqrt{\<\psi(t)|\psi(t)\>}}.
\end{equation}
Note again that this normalization procedure \textit{only} removes the non-norm-preserving aspect of non-unitary dynamics, with other impact of non-unitary dynamics still captured by the normalized state $|\phi(t)\>$. We then have
\begin{equation}
\<\phi(t)|\dot{\phi}(t)\> = \frac{\<\psi(t)|\dot{\psi}(t)\>}{\<\psi(t)|\psi(t)\>} - \frac{1}{2}\frac{\rd}{\rd t}\ln \<\psi(t)|\psi(t)\>,
\label{phit}
\end{equation}
where the overhead dot denotes the time derivative, $|\dot{\phi}(t)\> = \frac{\rd}{\rd t}|\phi(t)\>$. It is obvious that $\<\phi(t)|\dot{\phi}(t)\>$ is always purely imaginary because $\<\phi(t)|\phi(t)\>=1$.  Via the construction above, $|\phi(t)\>$ is also a cyclic state, with the following property
\begin{equation}
|\phi(T)\> = \e^{\ri\,\Re\alpha} |\phi(0)\>.
\end{equation}

Consider next the following time-evolving state
\begin{equation}
|\varphi(t)\> \equiv \e^{-\ri f(t)}|\phi(t)\>,
\end{equation}
with the real function $f(t)$ satisfying
\begin{equation}
f(T)-f(0)=\Re\alpha.
\end{equation}
Clearly,
\begin{equation}
|\varphi(T)\>= |\varphi(0)\>.
\end{equation}
So $|\varphi(t)\>$ is a periodic function of $t$, thus being a single-valued function along a closed curve in the projective Hilbert space traced out by $|\psi(t)\>$. A connection of $|\varphi(t)\>$ along this closed curve, namely, $\<\varphi(t)|\dot{\varphi}(t)\>$, can be well defined and easily evaluated. One immediately has
\begin{equation}
\<\varphi(t)|\dot{\varphi}(t)\> = -\ri \dot{f}(t) + \<\phi(t)|\dot{\phi}(t)\>.
\end{equation}
The AA phase is then obtained as an integral of the connection of $|\varphi(t)\>$ along the closed curve in the projective Hilbert space,
\begin{eqnarray}
\beta &\equiv& \ri \int_0^T \rd t\, \<\varphi(t)|\dot{\varphi}(t)\> \nn\\
&=& f(T) - f(0) + \ri \int_0^T \rd t\, \<\phi(t)|\dot{\phi}(t)\> \label{betaeq1}\\
&=& \frac{1}{\ri} \ln \frac{\<\psi(0)|\psi(T)\>}{\<\psi(0)|\psi(0)\>} + \ri \int_0^T \rd t\, \frac{\<\psi(t)|\dot{\psi}(t)\>}{\<\psi(t)|\psi(t)\>} \label{eq1-g} \\
&=&\alpha + \ri \int_0^T \rd t\, \frac{\<\psi(t)|\dot{\psi}(t)\>}{\<\psi(t)|\psi(t)\>}.
\label{eqn:newbeta}
\end{eqnarray}

A few important remarks are in order. First, because $\<\varphi(t)|\dot{\varphi}(t)\>$ is purely imaginary as $\<\phi(t)|\dot{\phi}(t)\>$, the AA phase $\beta$ obtained above is always real, irrespective of the context of the cyclic state $|\psi(t)\>$.  Second, $\beta$ is gauge-invariant \cite{Child97}.  That is, multiplying $|\psi(t)\>$ by an arbitrary time-dependent c-number factor, one obtains precisely the same AA phase.  This further confirms that the AA phase obtained above reflects the geometry of a closed curve in a projective Hilbert space. In addition, it can be also easily checked that $\beta$ can be understood as a consequence of a parallel transport along this curve. Third, if the cyclic dynamics is known to arise from adiabatic following of some instantaneous eigenstates of some (non-Hermitian) Hamiltonian (see below), then the AA phase will become the Berry phase by definition, and the resulting Berry phase must be always real, too.  Indeed, if the cyclic state follows instantaneous energy eigenstates, then they differ only by a dynamical phase factor and the gauge invariance guarantees that the AA phase is the same as the Berry phase. Fourth, the above final expression for AA phase is almost the same as in Ref.~\cite{Child97},  with the only difference being the $\<\psi(0)|\psi(0)\>$ factor in the first term of Eq.~(\ref{eq1-g}). That is, the earlier expression for AA phase~\cite{Child97} can equally apply to arbitrary non-unitary dynamics so long as $\<\psi(0)|\psi(0)\>=1$, with Eq.~(\ref{eq1-g}) reducing to Eq.~(22) in \cite{Child97}. This fourth point is one very interesting (side) finding here.  Interestingly, authors of Ref.~\cite{Child97} did not realize the generality and the always real nature of the $\beta$ expression above. Instead, they \cite{Child98} adopted an approach based on biorthonormal basis~\cite{GW88} to tackle non-unitary dynamics. That approach yields complex Berry connection and complex geometric phases in general \cite{GW88,Child98,Zheng15}, of which the physics is unclear as compared with the original meaning of geometric phase.

\subsection{Non-unitary dynamics in systems with non-Hermitian Hamiltonians}
If $|\psi(t)\>$ is governed by a Schr\"{o}dinger equation  with a non-Hermitian Hamiltonian $H(t)$,
\begin{equation}
\ri\hbar | \dot{\psi}(t)\> = H(t) | \psi(t)\>,
\label{eqn:Schr}
\end{equation}
then Eq.~(\ref{phit}) becomes
\begin{equation}
\<\phi(t)|\dot{\phi}(t)\> = \frac{1}{2\ri\hbar}\<\phi(t)|H(t) + H^\dag(t)|\phi(t)\>.
\end{equation}
The AA phase obtained in Eq.~(\ref{betaeq1}) becomes
\begin{eqnarray}
\label{beta2}
\beta
&=& \Re\alpha + \frac{1}{2\hbar} \int_0^T \rd t\, \<\phi(t)|H(t) + H^\dag(t)|\phi(t)\>\nn\\
&=& \Re\alpha + \frac{1}{\hbar}\Re \int_0^T \rd t\, \<\phi(t)|H(t)|\phi(t)\>\nn\\
&=& \Re\alpha + \frac{1}{\hbar} \Re \int_0^T \rd t\, \frac{\<\psi(t)|H(t)|\psi(t)\>}{\<\psi(t)|\psi(t)\>}.
\end{eqnarray}
Interestingly, plugging Eq.~(\ref{eqn:Schr}) into Eq.~(\ref{eqn:newbeta}), one arrives at an alternative but equivalent expression, namely,
\begin{eqnarray}
\label{beta3}
\beta= \alpha + \frac{1}{\hbar} \int_0^T \rd t\, \frac{\<\psi(t)|H(t)|\psi(t)\>}{\<\psi(t)|\psi(t)\>}.
\end{eqnarray}
However, it should be stressed that the AA phase expressions of Eqs.~(\ref{betaeq1}) and (\ref{eqn:newbeta}) are general and Eq.~(\ref{beta2}) and Eq.~(\ref{beta3}) only apply to those cases where the time evolution is governed by a Schr\"{o}dinger equation.  Related to this, it is also necessary to discuss the dynamical phase.  It is often said that the overall phase of a cyclic state is the sum of a dynamical phase and a geometric phase.  Applying this understanding to Eq.~(\ref{eqn:newbeta}) or Eq.~(\ref{beta3}), it is seen that $\alpha$ is the overall phase complex in general, whereas the dynamical phase $-\frac{1}{\hbar} \int_0^T \rd t\, \frac{\<\psi(t)|H(t)|\psi(t)\>}{\<\psi(t)|\psi(t)\>}$ is also complex in general. On the other hand, Eq.~(\ref{beta2}) indicates that, in representation of normalized states $|\phi(t)\>$, both the overall phase and the dynamical phase are always real. This second perspective is consistent with the one adopted by Samuel and Bhandari \cite{Samuel88}. It is important to note that the above two pictures based on $|\psi(t)\>$ and $|\phi(t)\>$ are equivalent because AA phase is gauge-invariant. The earlier criticisms in Ref.~\cite{Wu96} on Ref.~\cite{Samuel88} in the related literature are hence unfounded.

\section{A solvable Model with Adiabatic Following}
\label{sec:solvable}
In this section, we investigate a periodically driven, non-Hermitian Hamiltonian (in dimensionless units) treated previously \cite{GW88,Wu96,Zheng15} as our first model.   The Hamiltonian is given by
\begin{equation}
H_1(t) = \left(
\begin{array}{cc}
\epsilon & \e^{-\ri \omega t}\\
\e^{\ri \omega t}& -\epsilon
\end{array}
\right),
\label{eqn:Hamiltonian}
\end{equation}
where $\omega$ is real and $\epsilon$ is complex in general. The propagator $U(t)$ of this system can be calculated analytically.  An arbitrary eigenstate of the one-period (Floquet) operator $U(T)$ is called ``cyclic", because upon one period time evolution, it only gains a phase factor (which can be complex). The possible breakdown of adiabatic following can be examined by analyzing the time-dependent projection of one time-evolving cyclic state onto the instantaneous eigenstates of $H_1(t)$.  Using the AA phase as one simple quantitative characterization, it is of particular interest to inspect if the AA phase in the slow driving limit can approach the Berry phase.

The Floquet operator $U(T)$ in general has two eigenstates $|u^\pm\>$, which generate the following two cyclic states:
\begin{equation}
|F^\pm(t)\> \equiv U(t)|u^\pm\>
= \e^{\mp\ri\Omega t-\ri\half\omega t+\ri\gamma_\pm} \left(
\begin{array}{c}
\cos\left(\half\Theta_\pm\right)\\
\sin\left(\half\Theta_\pm\right)\e^{\ri\Phi_\pm}
\end{array}
\right),
\label{eqn:spherical}
\end{equation}
where
\begin{eqnarray}
\Omega &\equiv& \sqrt{1+\left(\epsilon-\half\omega\right)^2}, \\
\Theta_\pm &=& 2 \cot^{-1} \left|\epsilon -\half\omega \pm \Omega\right|, \label{thetadefine}\\
\Phi_\pm &=& \omega t -\gamma_\pm,
\end{eqnarray}
and
$\gamma_\pm$ is the phase of the complex variable $\left(\epsilon-\half\omega \pm \Omega\right)$.

The AA phase $\beta^\pm$ for $|F^\pm(t)\>$ is then found by plugging into Eq.~(\ref{eq1-g}). Specifically,
\begin{equation}
\beta^\pm = \frac{1}{\ri} \ln \frac{\<F^\pm(0)|F^\pm(T)\>}{\<F^\pm(0)|F^\pm(0)\>} + \ri \int_0^T \rd t\ \frac{\<F^\pm(t)|\dot{F}^\pm(t)\>}{\<F^\pm(t)|F^\pm(t)\>}.
\label{eq1-sec1}
\end{equation}
As elaborated in Sec.~\ref{sec:revisit}, the above expression for the AA phase is always real in arbitrary non-unitary dynamics. For the model in \Eqn{eqn:Hamiltonian}, we obtain
\begin{equation}
\beta^\pm = \frac{2\pi\left|\epsilon - \half\omega \pm \Omega\right|^2}{\left|\epsilon - \half\omega \pm \Omega\right|^2+1} .
\label{eqn:exactAAphase}
\end{equation}
Clearly, AA phases in Eq.~(\ref{eqn:exactAAphase}) are nothing but the half of the solid angles traced out by the cyclic states $|F^\pm(t)\>$ on the Bloch sphere, \ie,
\begin{equation}
\beta^\pm=\pi\left(1+\cos\Theta_\pm\right),
\end{equation}
where $\Theta_\pm$ is given in Eq.~(\ref{thetadefine}).  This explicitly confirms the physical meaning of the AA phase adopted in this work.

We are now ready to examine what happens in the slow driving limit, namely, cases with $\omega\to0$. First of all, the AA phases in \Eqn{eqn:exactAAphase} reduce to
\begin{equation}
\beta^\pm \to  \frac{2\pi\left|\epsilon \pm \sqrt{1+\epsilon^2}\right|^2}{\left|\epsilon \pm \sqrt{1+\epsilon^2}\right|^2+1}.
\label{eqn:AA-Berry}
\end{equation}
On the other hand, cyclic states $|F^\pm(t)\>$ are found to reduce to instantaneous energy eigenstates of $H_1(t)$ up to overall phases. This indicates that $\beta^\pm$ obtained above become Berry phases (they are certainly real). Interestingly, the Berry phase obtained in Ref.~\cite{GW88} in our notation would be $ \beta^-_{\rm GW} = \pi\left( 1- \frac{\epsilon}{\sqrt{1+\epsilon^2}}\right)$, which is complex in general and does not have a simple geometrical interpretation as the original Berry phase.

\section{Computational studies of a ``hopping'' model}
\label{sec:hopping}
Here we computationally study a second simple model with the following non-Hermitian Hamiltonian (in dimensionless units)
\begin{equation}
H_2(t) =  \left(
\begin{array}{cc}
1 & \ri \mu (\cos\omega t+\ri)\\
\ri \mu (\cos\omega t+\ri)& -1
\end{array}
\right).
\label{eqn:hopping}
\end{equation}
As previously shown by us \cite{GongWang15}, though $H_2(t)$ is non-Hermitian, its time evolution can be stable because its cyclic states may only acquire a real overall phase after one period.  Our results discussed below are indeed in this stable region. Analytical cyclic states here are not available. We hence use two numerically obtained cyclic states as the initial states. We then analyze the details of the resulting cyclic time evolution in terms of the ratio of the two components of the time-evolving state, as compared with the instantaneous eigenstates of $H_2(t)$. Of particular interest here is what happens if the driving period $T=2\pi/\omega$ becomes large.

\begin{figure}[h!]
	\begin{center}
		$\begin{array}{cc}
		\qquad\qquad~~{\rm (a)}& \qquad\quad{\rm (b)}\\
		\includegraphics[width=0.48\columnwidth]{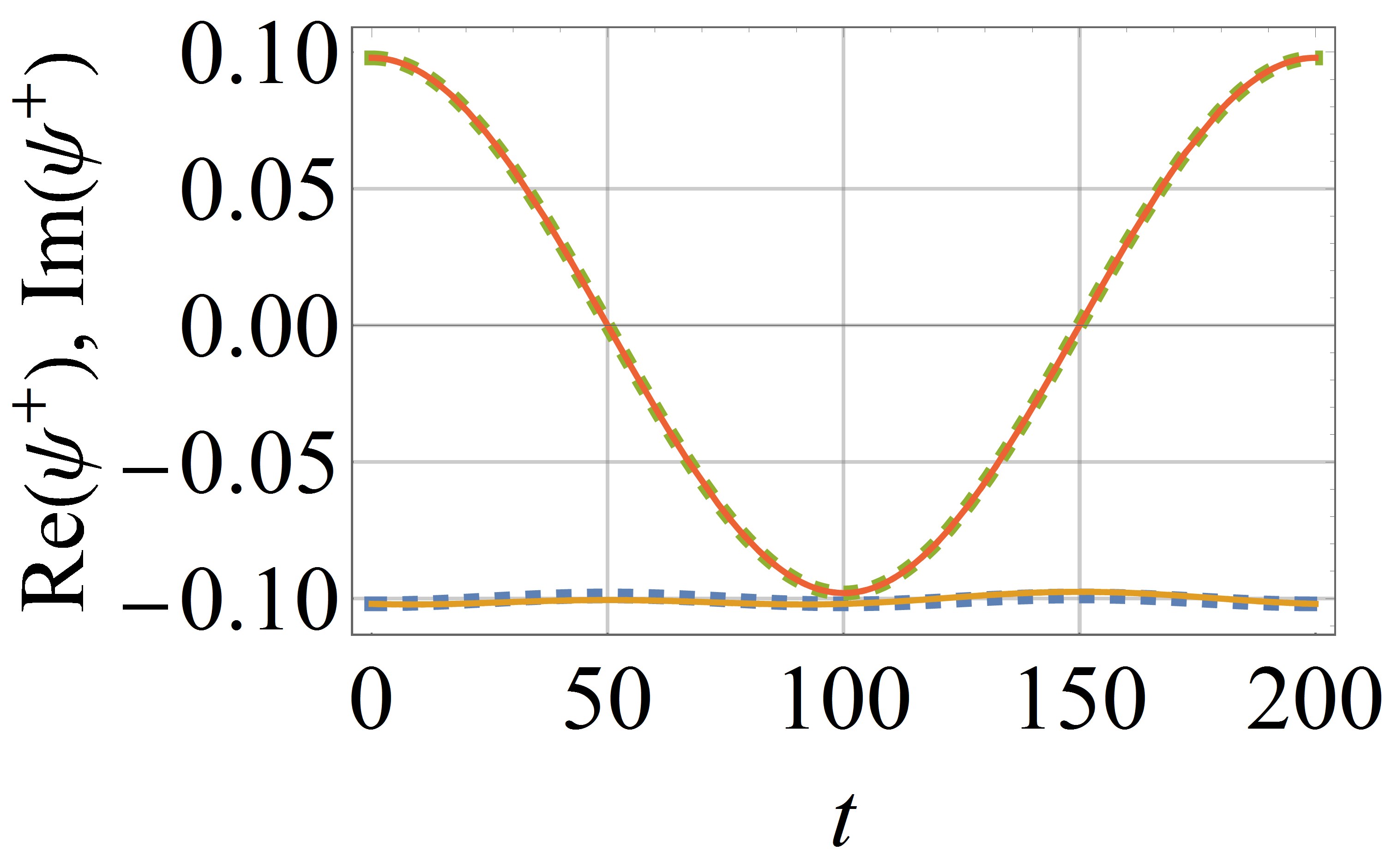}&
		\includegraphics[width=0.48\columnwidth]{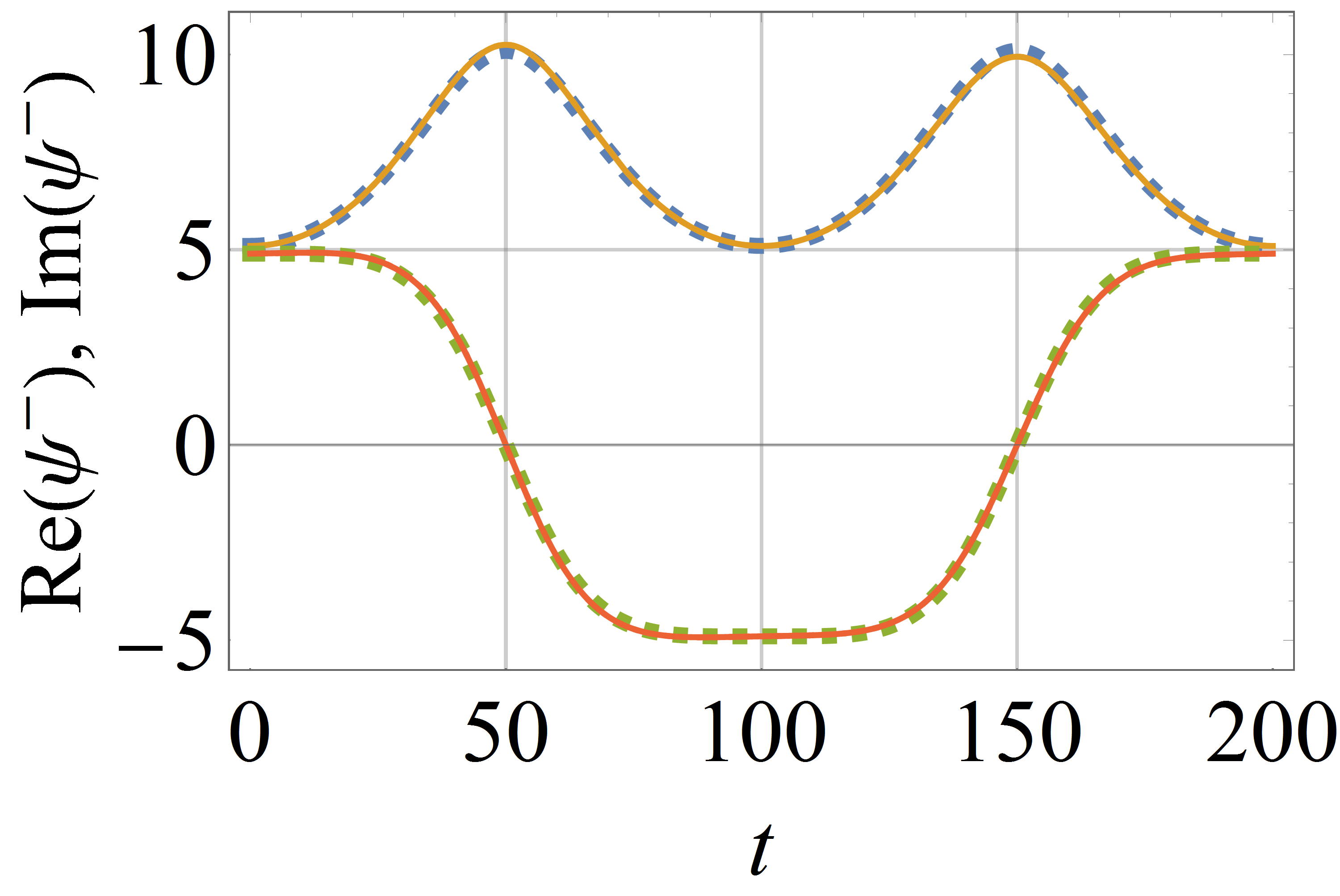}
		\end{array}$
		\caption{\label{fig:following}(color online) Comparison between the cyclic states of the model Eq.~(\ref{eqn:hopping}) for $\mu=0.2$ and the instantaneous eigenstates of $H_2(t)$ for sufficiently slow driving $(T=200)$. For a state $[a(t), b(t)]^T$, the plotted vertical coordinates represent the time dependence of the real and imaginary parts of the ratios [denoted $\psi=b(t)/a(t)$] of the two components of the time-evolving cyclic states (solid lines) or that of the eigenstates of $H_2(t)$ (dotted line).  Panels (a) and (b) are for two different cyclic states. Note that the solid lines almost perfectly overlap with dotted lines indicates adiabatic following.}
	\end{center}
\end{figure}

For small values of $\mu$, it is found that the cyclic states can follow the instantaneous eigenstates of $H_2(t)$ as the driving slows down. As illustrated in Fig.~\ref{fig:following},  for $T$ of the order of hundreds, adiabatic following of the cyclic states with the $H_2(t)$ eigenstates is already clearly visible. Indeed, the AA phase obtained numerically approaches a zero geometric phase, in agreement with a direct calculation [based on instantaneous eigenstates of $H_2(t)$] that gives a zero Berry phase.  All these features are analogous to what we obtain in Sec.~\ref{sec:solvable}.

\begin{figure}[h!]
	\begin{center}
		$\begin{array}{cc}
		\qquad~~~{\rm (a)}& \qquad~~~{\rm (b)}\\
		\includegraphics[width=0.48\columnwidth]{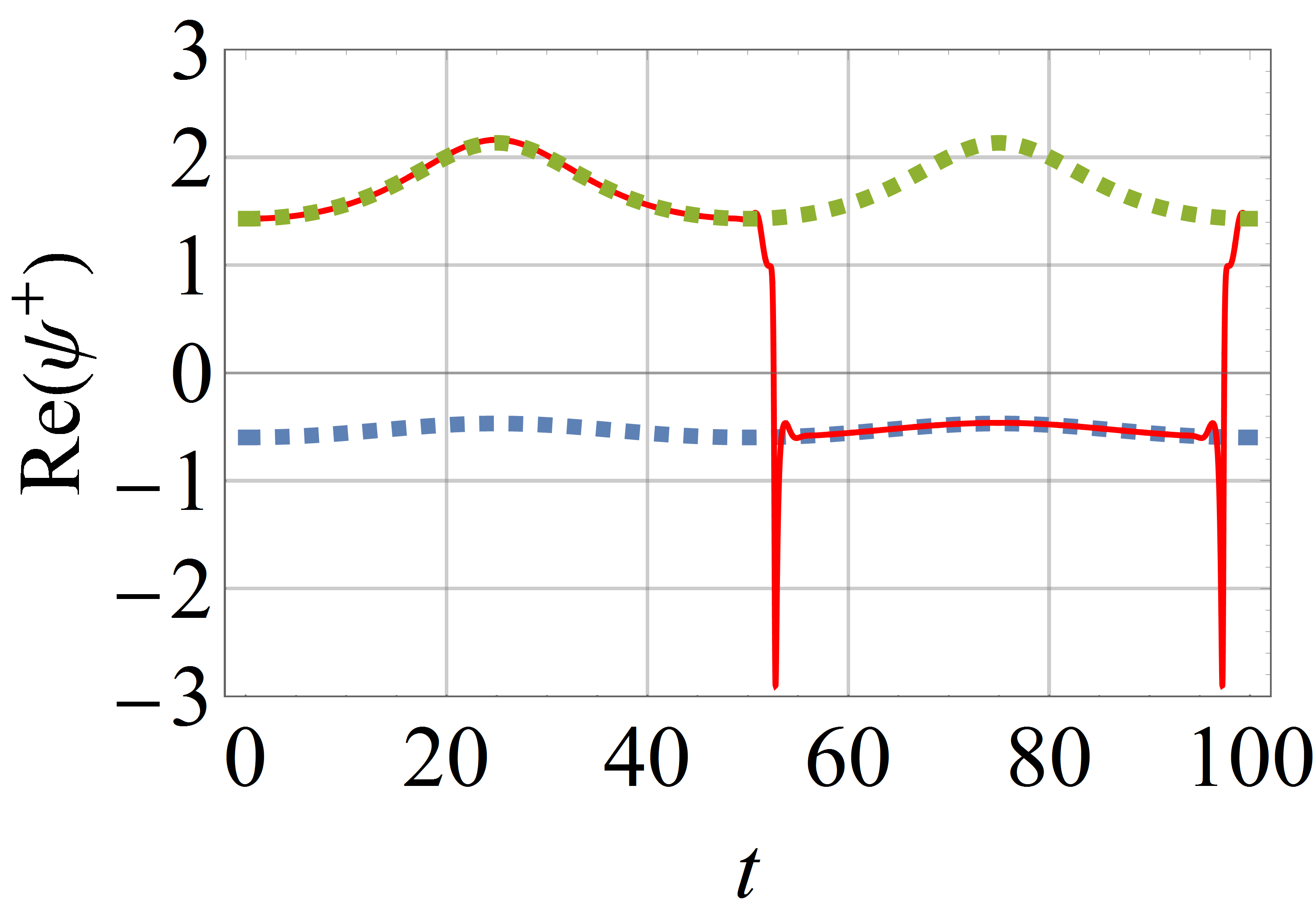}&
		\includegraphics[width=0.48\columnwidth]{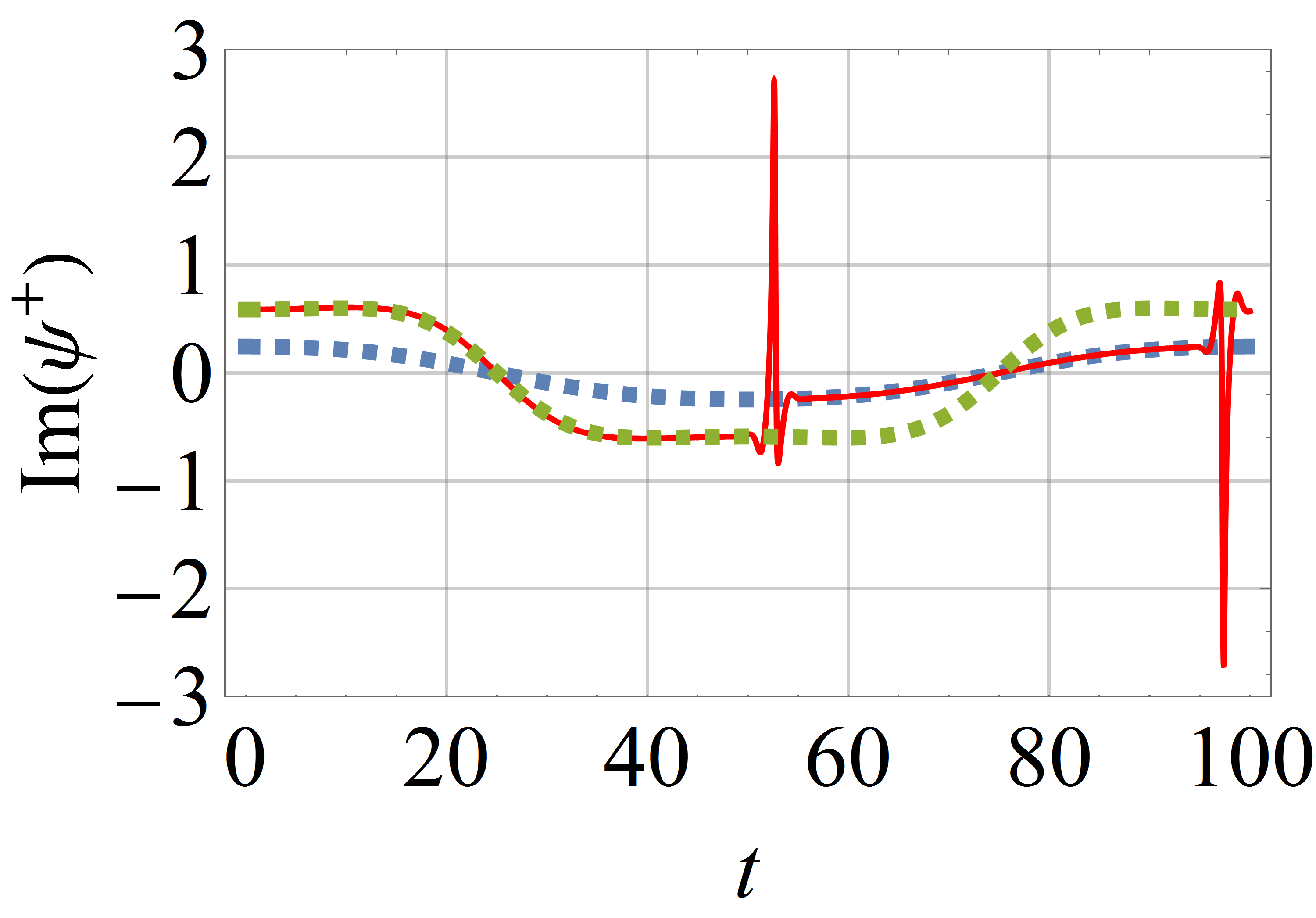}\\
		\qquad~~~{\rm (c)}& \qquad~~~{\rm (d)}\\
		\includegraphics[width=0.48\columnwidth]{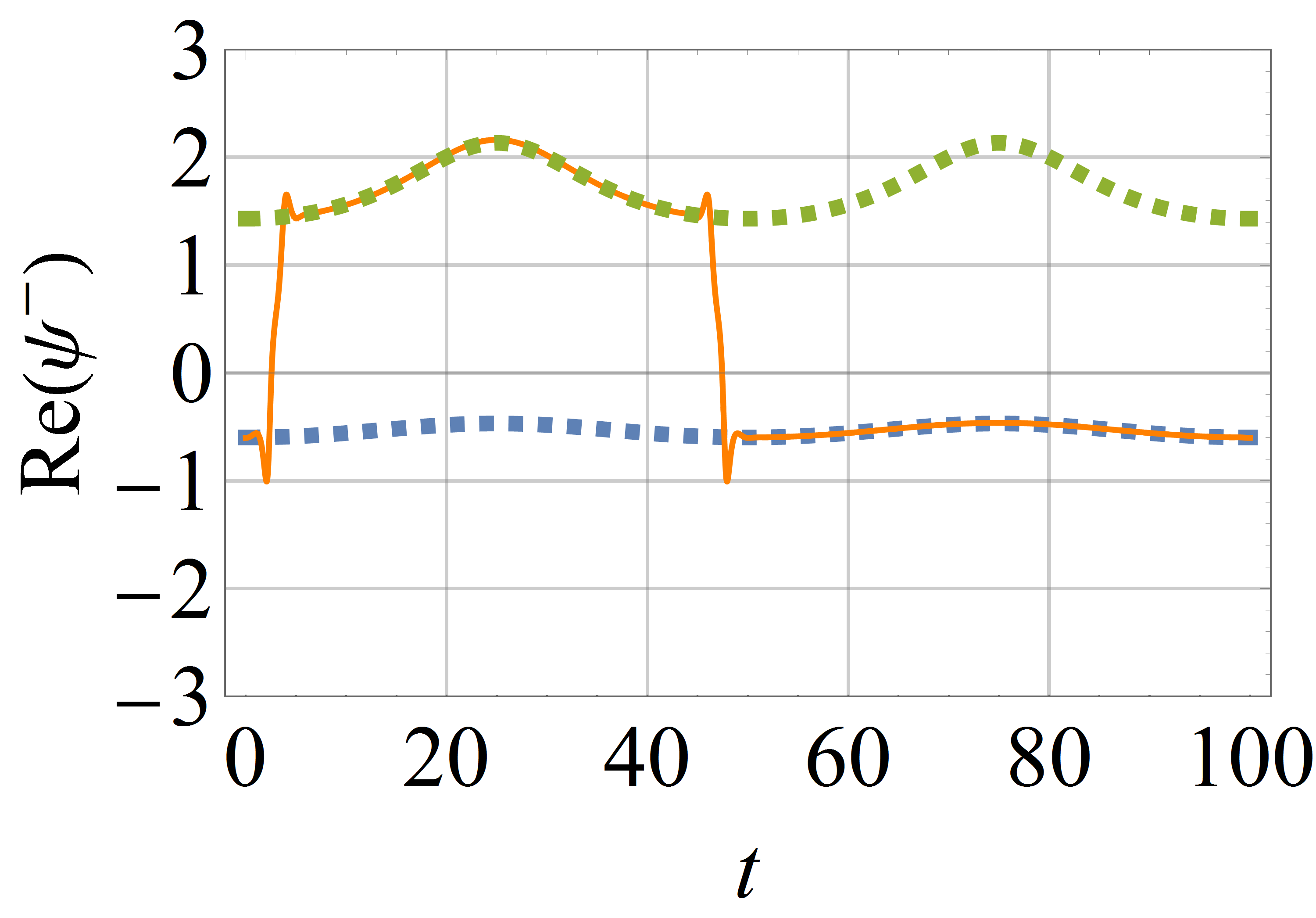}&
		\includegraphics[width=0.48\columnwidth]{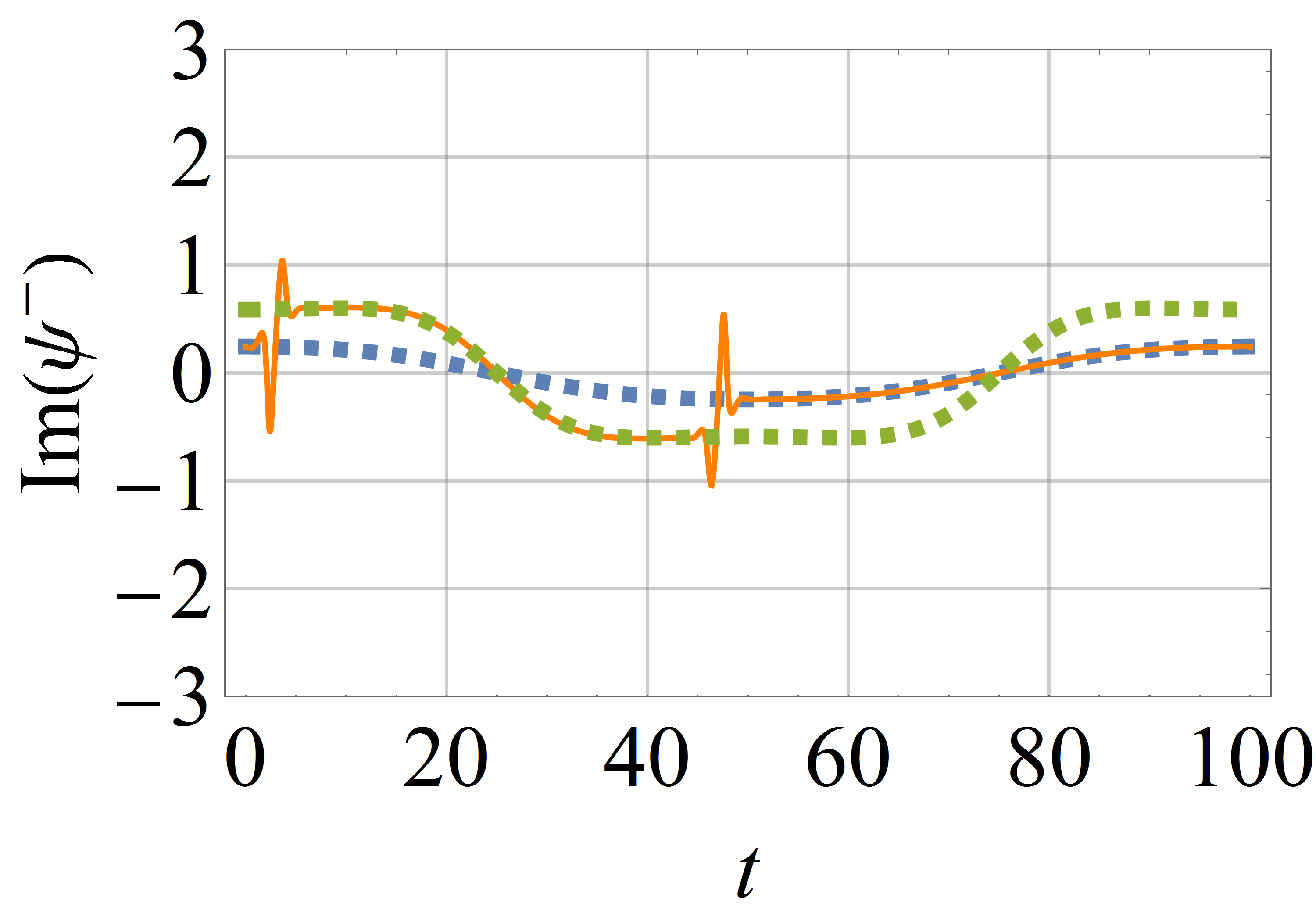}
		\end{array}$
		\caption{\label{fig:hopping}(color online)
			Comparison between two cyclic states of the model depicted by Eq.~(\ref{eqn:hopping}) for $\mu=1.2$ and the two instantaneous eigenstates of $H_2(t)$ for sufficiently slow driving $(T=100)$. For a state $[a(t), b(t)]^T$, the plotted vertical coordinates represent the ratios [denoted $\psi=b(t)/a(t)$] of the two components of time-evolving cyclic states (solid lines), as compared with the parallel behavior of two instantaneous eigenstates of $H_2(t)$ (upper and lower dotted lines). Panels (a) and (b) are for one cyclic state, and panels (c) and (d) are for the other cyclic state. Note that both cyclic states here exhibit hopping.}
	\end{center}
\end{figure}

However, as shown in Fig. 2, for larger values of $\mu$, the above observations break down completely. In cases of slow driving, the time evolution of the cyclic states now displays exotic dynamics by {\it hopping} between two instantaneous eigenstates. Before and after one hopping, a cyclic state tends to follow one of the instantaneous eigenstates of $H_2(t)$. That is, the adiabatic following is only true \textit{piecewise}.  This clearly shows that when the overall time evolution is stable, local instability can still dominate over the dynamics during certain time windows that can be very small compared with the driving period.  It should be noted that the hopping behavior observed here is displayed by a time-evolving state projected onto well-behaved instantaneous eigenstates [see, e.g. Eq.~(29)].   It is hence unrelated to any numerical instabilities when parametrically tracing the instantaneous eigenstates of a non-Hermitian system \cite{newRef2}.

The hopping phenomenon here thus demonstrates that adiabatic following in non-unitary dynamics may not hold.  We now discuss one important difference between our finding and previous studies.  In Refs.~\cite{R2,Rotter}, one typically  uses  an instantaneous eigenstate of the Hamiltonian as the initial state, and then adiabatic following is observed to break down due to non-negligible accumulation of non-adiabatic transitions in the non-unitary dynamics.  By contrast, here we instead use a cyclic state as the initial state (which is very close to, but not the same as instantaneous eigenstates of the Hamiltonian for slow driving). Now even though the time evolution is both stable and cyclic, the dynamics still displays intriguing hopping and hence violates adiabatic following.  This hints that the breakdown of adiabatic following observed here is on a different (and perhaps more fundamental) level than studied before \cite{Rotter}. Indeed, if we also use cyclic states as the initial states for the model studied in Ref.~\cite{Rotter}, then we still obtain perfect adiabatic following under slow driving. In addition, as elaborated in Sec.~\ref{sec:BU}, the hopping dynamics found here has a root in a genuine phase transition in the parameter space.

\begin{figure}[h!]
	\begin{center}
		\includegraphics[width=0.7\columnwidth]{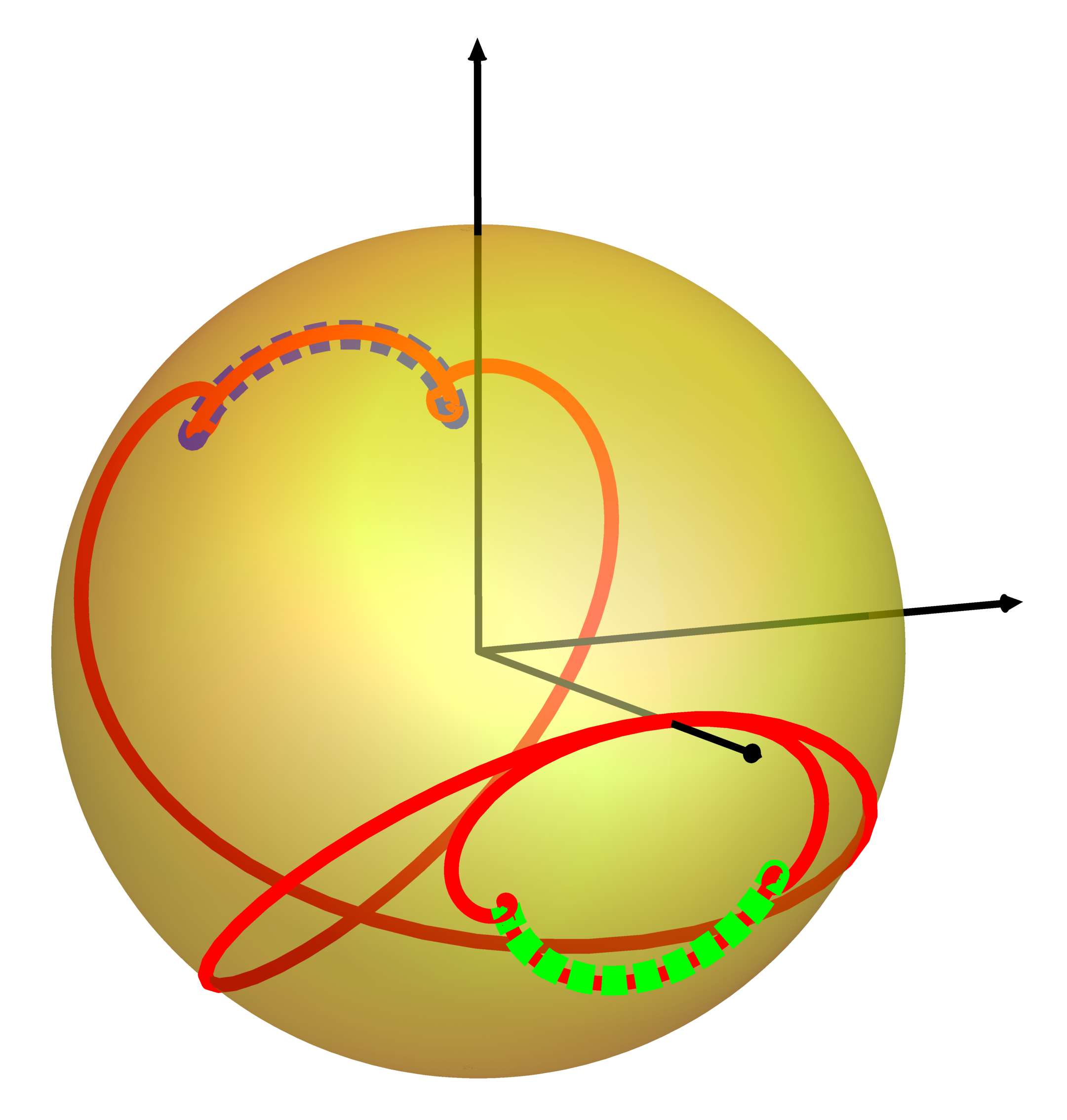}
		\caption{\label{fig:sphere}(color online)
			Geometry of one cyclic state (solid line) plotted on the Bloch sphere (one of the two considered in Fig.~\ref{fig:hopping} with $\mu=1.2$), as compared with the geometry of two instantaneous eigenstates of $H_2(t)$ (dotted lines).}
	\end{center}
\end{figure}

Figure \ref{fig:sphere} presents on the Bloch sphere the exotic {hopping} dynamics of one cyclic state shown in Fig.~\ref{fig:hopping}. It is seen that,  due to the {hopping} between two instantaneous eigenstates, the geometry of the curve traced out by the cyclic state becomes highly nontrivial. Specifically, in this example each instantaneous eigenstate of $H_2(t)$ does not trace out a solid angle on the Bloch sphere (indicating zero Berry phase), but the cyclic state does trace out (via hopping) a significant solid angle on the Bloch sphere, thus yielding a nonzero AA phase.

\begin{figure}[h!]
	\begin{center}
		\includegraphics[width=0.9\columnwidth]{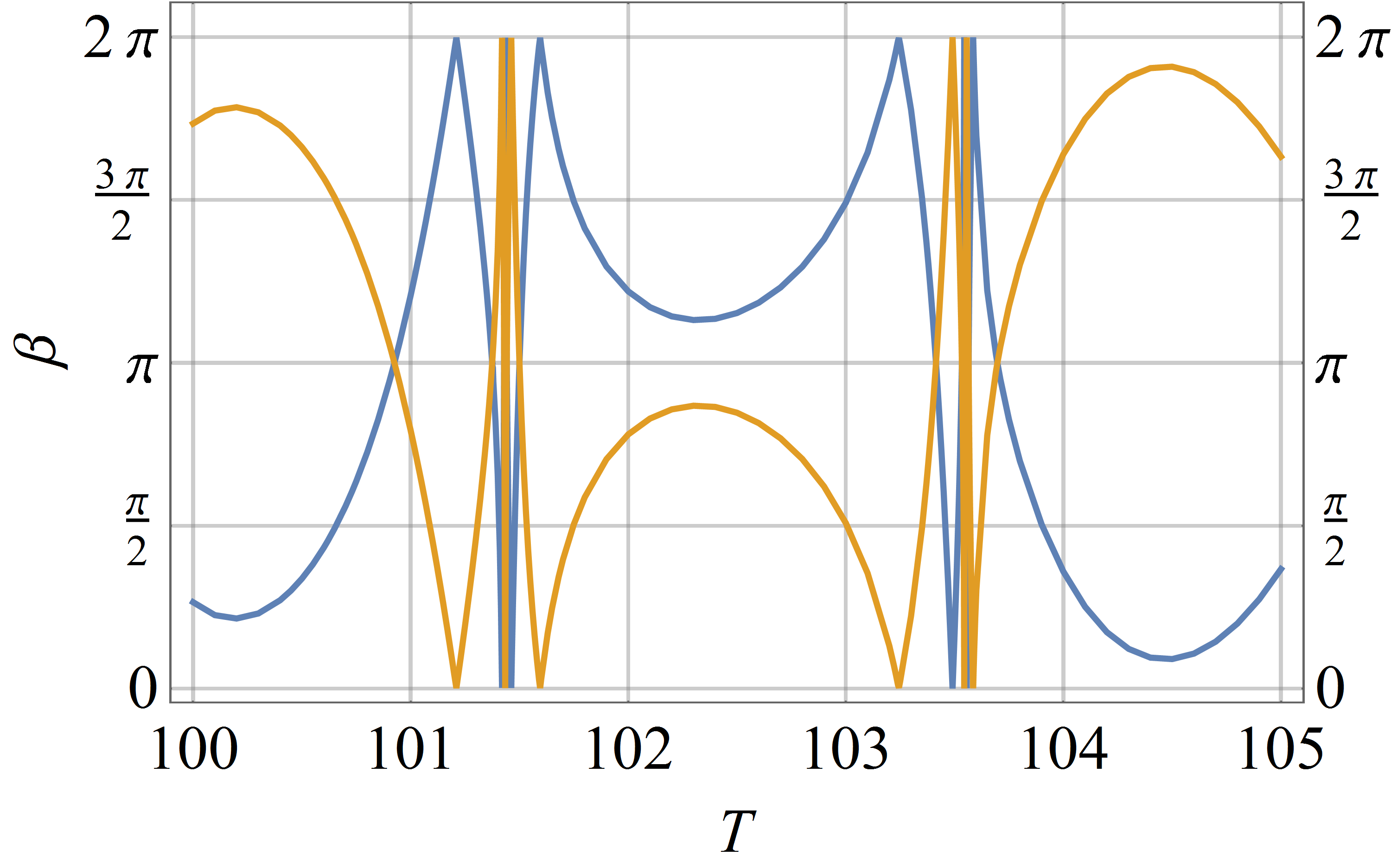}
		\caption{\label{fig:beta}(color online) AA phases $\beta$ as functions of period $T$ in the hopping model $H_2(t)$ with $\mu=1.2.$ Note the small range in $T$ and the violent oscillations in $\beta$.}
	\end{center}
\end{figure}

We further look into the sensitivity of the obtained AA phase to the exact values of $T$.  We find that due to the {hopping} behavior of the cyclic states, the actual geometry of a cyclic state changes drastically as $T$ is tuned. The resulting AA phase in general does not approach any limit. For example, Fig.~\ref{fig:beta} presents the AA phases vs $T$ for the two cyclic states considered in Fig.~\ref{fig:hopping}, for a {\it small} time window $T$. It is seen that the AA phase for each individual cyclic state can be extremely sensitive to $T$, and oscillate violently between $0$ and $2\pi$.  Based on these observations and other calculations not shown here, we conclude that the AA phase in the hopping model cannot have a large-$T$ limit. The high sensitivity of the AA phase to $T$ hints that the geometry of cyclic states in non-unitary dynamics is extremely rich.

\section{Berry-Uzdin Model}
\label{sec:BU}
\subsection{The Berry-Uzdin model}
To develop theoretical insights into the hopping phenomenon or piecewise adiabatic following observed in Sec.~\ref{sec:hopping}, we revisit the Berry-Uzdin model \cite{BU},
\begin{equation}
H_{\rm BU} (t) = \ri\left(
\begin{array}{cc}
0&1\\
z[\theta(t)]&0
\end{array}
\right),
\label{eqn:BU}
\end{equation}
with
\begin{equation}
z[\theta(t)] \equiv \rho \e^{\ri\theta(t)} - r,
\end{equation}
where $\rho$ and $r$ are real variables. Because $z(\theta+2\pi) = z(\theta)$, this is hence another periodically driven model. The instantaneous  eigenvalues of $H_{\rm BU}$ are given by
\begin{equation}
E_\pm(t) = \pm \ri \sqrt{z[\theta(t)]},
\end{equation}
with the corresponding eigenstates found to be
\begin{equation}
|E_\pm(t)\> = \left(
\begin{array}{c}
z^{-1/4}[\theta(t)]\\
\pm z^{1/4}[\theta(t)]
\end{array}\right).
\end{equation}
They form a complete set as long as the eigenvalues are not degenerate.

The Berry-Uzdin model depicted above is exactly solvable if $\theta$ is a linear function of time,
\begin{equation}
\theta(t)= 2\pi \frac{t}{T}.
\end{equation}
From now on, $\theta(t)=2\pi \frac{t}{T}$ can be regarded as a rescaled time variable. We then find the Floquet eigenstates (cyclic states) as below, whose explicit time dependence is given by,
\begin{equation}
|F^\pm(t)\> = \left(
\begin{array}{c}
J_{\pm\nu}[\nu x(\theta)]\\
\p_t J_{\pm\nu}[\nu x(\theta)]
\end{array}\right),
\end{equation}
where
\begin{equation}
\nu \equiv\frac{T}{\pi}\sqrt{r}, \qquad
x(\theta) \equiv \sqrt{\frac{\rho}{r}} \e^{\ri\theta/2},
\end{equation}
and $J_{\pm\nu}$ being the Bessel functions.  Noticing that
\begin{equation}
|F^\pm(T)\> = \e^{\pm\ri T\sqrt{r}} |F^\pm(0)\>,
\end{equation}
one finds that this periodically driven system is stable (that is, having extended unitarity) for $r>0$ and unstable for $r<0$ \cite{Joglekar,GongWang15}.  We also note that though the Berry-Uzdin model was studied before, its dynamics with one cyclic state as the initial state was not previously paid attention to.

\subsection{Asymptotic analysis}

To develop an explicit theory to account for a hopping mechanism, we expand the Floquet states $|F^\pm(T)\>$ in terms of the instantaneous eigenstates of $H_{\rm BU} (t)$,
\begin{equation}
|F^\pm(t)\> = a ^\pm_+(t) |E_+(t)\> + a ^\pm_-(t) |E_-(t)\>,
\end{equation}
with the expansion coefficients given by
\begin{eqnarray}
a^\pm_+(t) &\equiv& \half z^{1/4}(\theta) J_{\pm\nu}(\nu x) + \half z^{-1/4}(\theta)\p_t J_{\pm\nu}(\nu x), \\
a^\pm_-(t) &\equiv& \half z^{1/4}(\theta) J_{\pm\nu}(\nu x) - \half z^{-1/4}(\theta)\p_t J_{\pm\nu}(\nu x).
\end{eqnarray}
For each one of the two cyclic states, we characterize the relative weightage of the two projection amplitudes with respect to $|E_\pm(t)\>$ via their ratio, namely,
\begin{eqnarray}
R^+(t) &\equiv& \frac{a^+_-}{a^+_+} = \frac{\sqrt{z(\theta)}J_\nu(\nu x) - \p_t J_\nu(\nu x)}{\sqrt{z(\theta)}J_\nu(\nu x) + \p_t J_\nu(\nu x)}, \\
R^-(t) &\equiv& \frac{a^-_+}{a^-_-} = \frac{\sqrt{z(\theta)}J_{-\nu}(\nu x) + \p_t J_{-\nu}(\nu x)}{\sqrt{z(\theta)}J_{-\nu}(\nu x) - \p_t J_{-\nu}(\nu x)}.
\end{eqnarray}
If $|R^-(t)|\ll 1$, it is safe to say that the cyclic state $|F^-(t)\>$ is following the instantaneous eigenstate $|E_-(t)\>$. If $|R^-(t)|\gg 1$, the cyclic state $|F^-(t)\>$ is instead following the instantaneous eigenstate $|E_+(t)\>$. The transition from $|R^-(t)|\ll 1$ to $|R^-(t)|\gg 1$, if completed within a small time window, then signifies a hopping behavior. Similar analysis can be applied to the other Floquet state $|F^+(t)\>$ using $|R^+(t)|$.

We focus on the case $\rho<r$, the so-called ``degeneracy-excluding loops'' in Ref.~\cite{BU}. In this case, $|x|<1$. For fixed and non-vanishing $r$ and $\rho$, the large $T$ limit becomes the same as the large $\nu$ limit. As $\nu \to \infty$, we have \cite{DLMF},
\begin{eqnarray}
J_\nu(\nu x) & \sim & \frac{\e^{-\nu \frac{2}{3} \xi^{3/2} }}{\sqrt{2\pi\nu}} \frac{1}{\left(1-x^2\right)^{1/4}}
, \\
J'_\nu(\nu x) & \sim & \frac{\e^{-\nu \frac{2}{3} \xi^{3/2} }}{\sqrt{2\pi\nu}} \frac{\left(1-x^2\right)^{1/4}}{x}
,
\end{eqnarray}
where we define a new variable $\xi$ through
\begin{eqnarray}
\frac{2}{3} \xi^ {3/2}   &= &
\ln \frac{1 + \sqrt{1-x^2}}{x} - \sqrt{1-x^2}\nn\\
&=& \ln \frac{\sqrt{r} + \sqrt{r-\rho \e^{\ri\theta}}}{\sqrt{\rho} \e^{\ri\theta/2}} - \frac{\sqrt{r-\rho \e^{\ri\theta}}}{\sqrt{r}}.
\end{eqnarray}
Clearly then, $\xi$ inherits its time-dependence from $\theta$. Plugging these intermediate results into the expression for $R^+$, one arrives at
\begin{equation}
R^+(t) \sim \frac{x^2}{4\nu(1-x^2)^{3/2}} = \frac{\pi}{4T} \frac{\rho\e^{\ri\theta}}{\left(r-\rho\e^{\ri\theta}\right)^{3/2}}, \quad T\to \infty.
\end{equation}
Because this  $|R^+(t)|$ (in the regime of $\rho<r$) smoothly approaches zero in the slow driving limit, we infer that there is always adiabatic following for the ``$+$'' Floquet state. Nevertheless, it is entirely a different story for the ``$-$'' Floquet state. As $\nu\to\infty$, we obtain  \cite{DLMF},
\begin{eqnarray}
J_{-\nu}(\nu x) & \sim & \frac{1}{\sqrt{2\pi\nu}\left(1-x^2\right)^{1/4}} \left[\cos(\nu\pi)\,\e^{-\nu \frac{2}{3} \xi^{3/2} }\right.\nn\\
	&&\quad \left.+ 2 \sin(\nu\pi)\, \e^{\nu \frac{2}{3} \xi^{3/2}} \right], \\
J'_{-\nu}(\nu x) & \sim & \frac{\left(1-x^2\right)^{1/4}}{x\sqrt{2\pi\nu}}  \left[\cos(\nu\pi)\, \e^{-\nu \frac{2}{3} \xi^{3/2} }\right.\nn\\
	&&\quad\left.- 2 \sin(\nu\pi)\, \e^{\nu \frac{2}{3} \xi^{3/2}} \right] .
\end{eqnarray}
Depending on the sign of the real part of the exponent $\nu \frac{2}{3} \xi^{3/2}$, one exponential will be dominant over the other exponential. This is nothing but the Stokes phenomenon. As a consequence, $R^-$ has two distinct types of behavior in two Stokes wedges. That is,
\begin{eqnarray}
R^-(t) &\sim&
\left\{
\begin{array}{ll}
\dsp\frac{x^2}{4\nu(1-x^2)^{3/2}},  &  \quad\Re  \xi^{3/2}> 0\\
\dsp-\frac{4\nu(1-x^2)^{3/2}}{x^2},  &  \quad\Re \xi^{3/2} < 0
\end{array}
\right.\nn\\
&=& \left\{
\begin{array}{ll}
\dsp\frac{\pi}{4T} \frac{\rho\e^{\ri\theta}}{\left(r-\rho\e^{\ri\theta}\right)^{3/2}}, &  \Re  \xi^{3/2}> 0\\
\dsp - \frac{4T}{\pi} \frac{\left(r-\rho\e^{\ri\theta}\right)^{3/2}}{\rho\e^{\ri\theta}}, &  \Re \xi^{3/2} < 0
\end{array}
\right.
\end{eqnarray}
Asymptotic analysis above shows the following:  In the slow driving limit $T\to\infty$, the Floquet state $|F^-(t)\>$ follows $|E_-(t)\>$ when $\Re  \xi^{3/2}> 0$ and it follows $|E_+(t)\>$ when $\Re  \xi^{3/2} < 0$. When $\Re  \xi^{3/2} $ flips its sign, the ``$-$'' Floquet state makes a switch between the two instantaneous eigenstates $|E_\pm(t)\>$.

\begin{figure}[h!]
	\begin{center}
		\includegraphics[width=0.9\columnwidth]{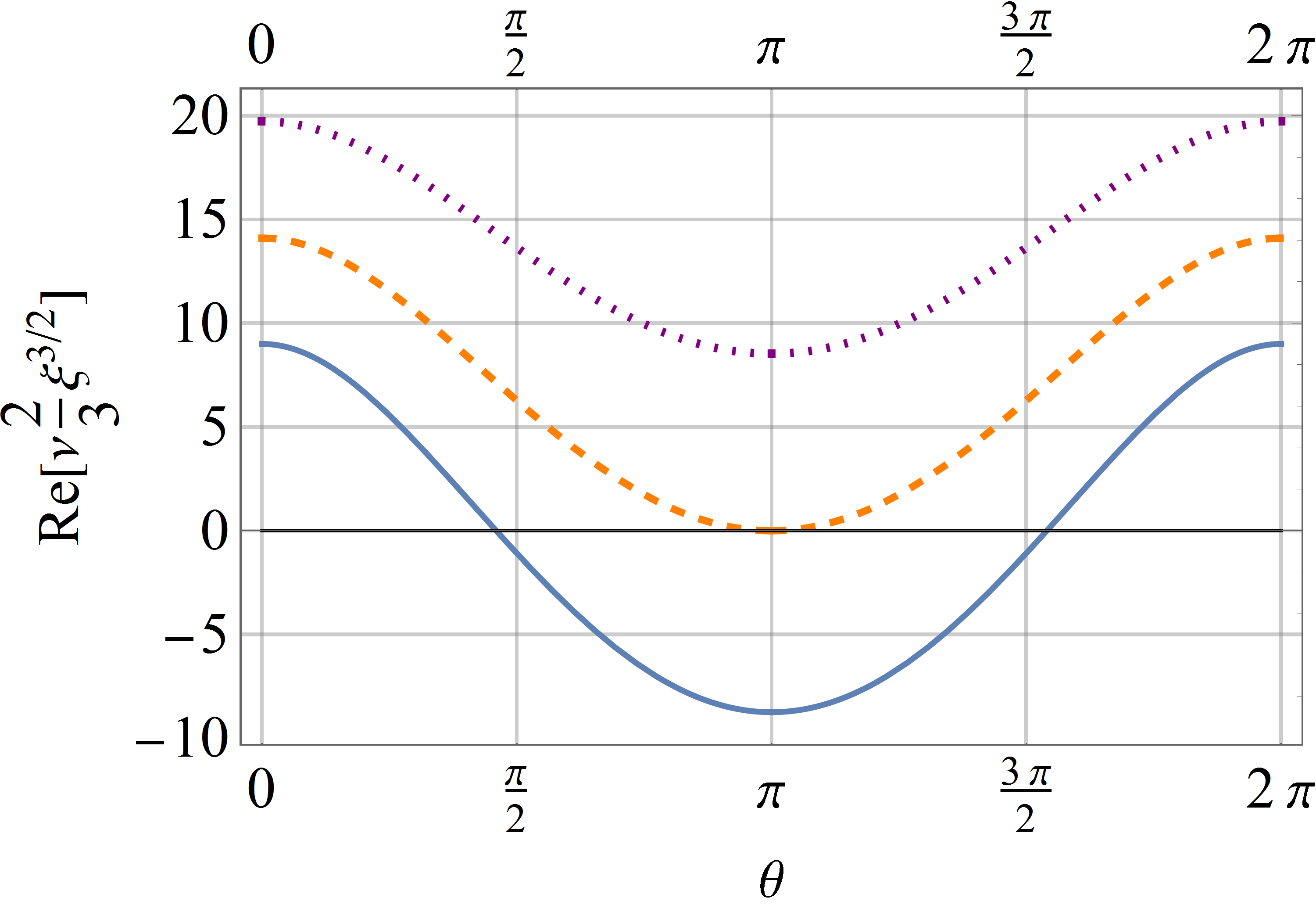}
		\caption{\label{fig:exponent}(color online) The real part of the exponents in the asymptotic behaviours of Bessel functions. The (blue) line is the exponent with $\rho/r=0.55$, the (orange) dashed line is for $\rho/r=c\approx0.439\,229$, and the (purple) dotted line is for $\rho/r=0.35$. For all the shown cases, $r=1$ and $T=200$.
		}
	\end{center}
\end{figure}

Still under our early assumption $\rho<r$ (the degeneracy excluding regime), we further obtain a critical value of $\rho$ for the hopping to occur.  The critical value can be located  by the condition $\Re  \xi^{3/2}=0$ at $\theta=\pi$. As illustrated in Fig.~\ref{fig:exponent},  one can find the following critical boundary
\begin{equation}
\left(\frac{\rho}{r}\right) _ \text{crit} = c \approx 0.439\,229\cdots,
\end{equation}
where the $c$ value is determined by the algebraic equation,
\begin{equation}
\ln \frac{1 + \sqrt{1+ c}}{\sqrt{c}} = \sqrt{1+c}.
\end{equation}

Two more explicit results can be obtained. Firstly,  if $\rho$ is slightly beyond the critical ratio, with $|\rho/r-c|\ll 1$, then we can estimate that the hopping occurs when
\begin{equation}
\theta\sim \pi \pm 2 \frac{\sqrt{1+c}}{c} \sqrt{\left|\frac{\rho}{r} - c\right|}  + \mathcal{O}\left(\frac{\rho}{r} - c\right) .
\end{equation}
That is, For $\rho/r$ slightly above its critical value, the hopping is predicted to occur near $\theta=\pi$. Secondly, for $c<\rho/r<1$, it is easy to find that the hopping should more or less complete within a time window given by
\begin{equation}
\Re \frac{2}{3} \xi^ {3/2} \sim \frac{1}{\nu} = \frac{\pi}{T\sqrt{r}}.
\end{equation}
This is a very interesting understanding.  The width of the hopping window in terms of $\theta$ is inversely proportional to the driving period $T$.  This finally justifies the terminology \textit{hopping} we advocate here: as we learn from this analytically solvable model, in the slow driving (large $T$) limit,  the time window needed for switching from following one instantaneous eigenstate to following the other one becomes smaller and smaller (as compared with the driving period). Note also that $\theta \propto t/T$, hence the absolute time needed for a complete hopping is found to be independent of $T$.

\subsection{Phase transition in cyclic dynamics}

\begin{figure}[h!]
	\begin{center}
		$\begin{array}{cc}
		\qquad\quad~~~{\rm (a)}& \qquad~~{\rm (b)}\\
		\includegraphics[width=0.48\columnwidth]{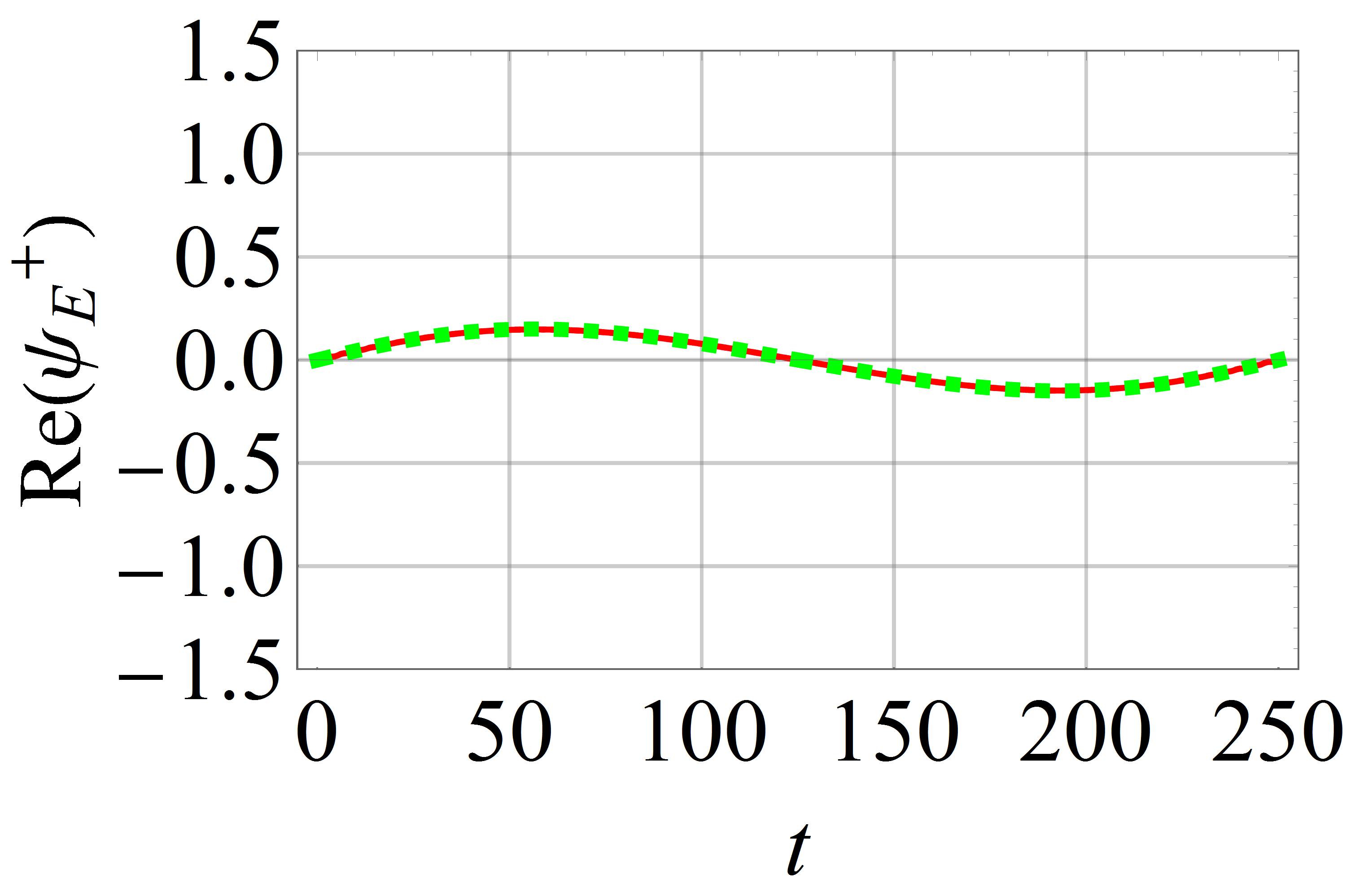}&
		\includegraphics[width=0.48\columnwidth]{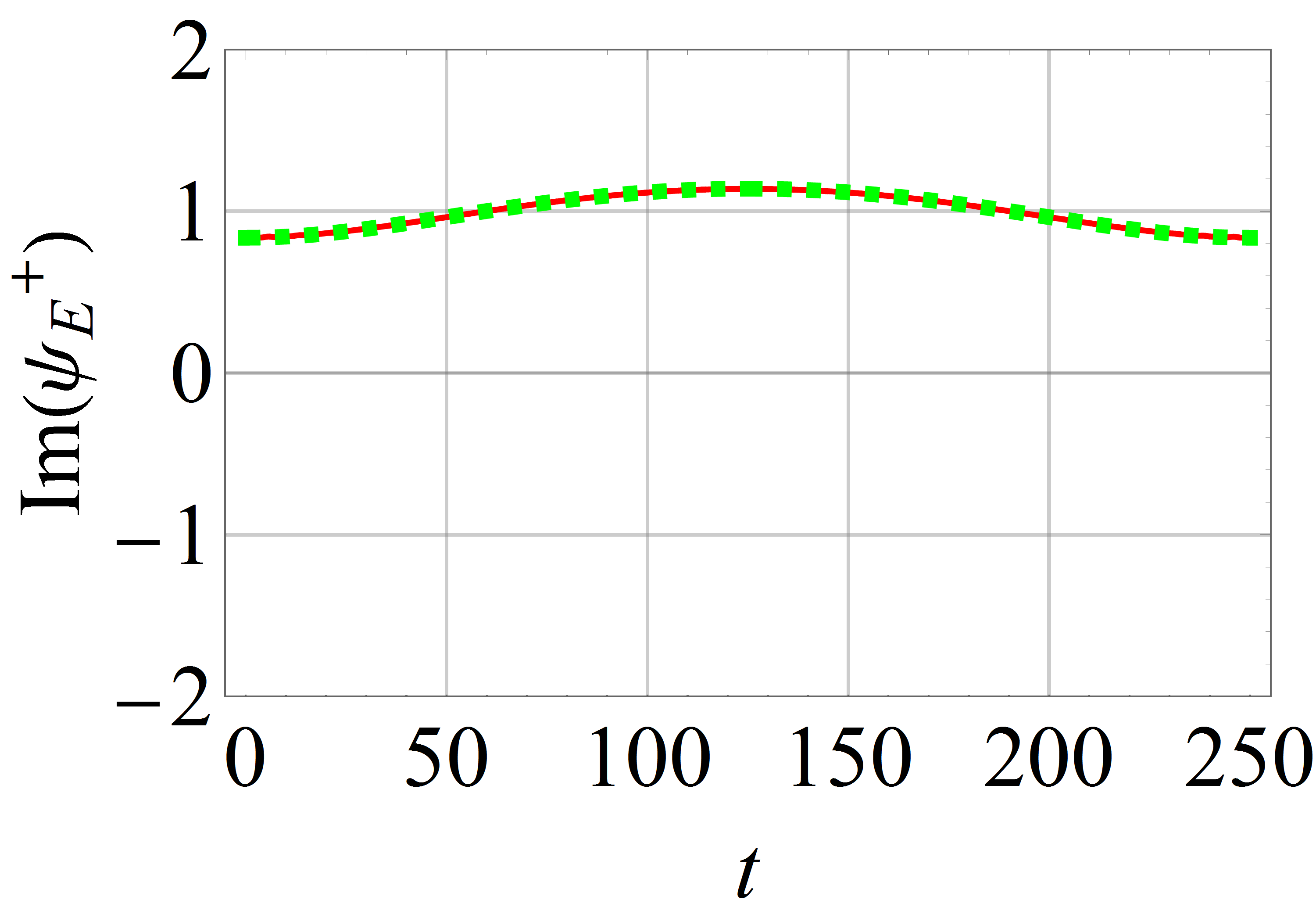}\\
		\qquad\quad~~~{\rm (c)}& \qquad~~{\rm (d)}\\
		\includegraphics[width=0.48\columnwidth]{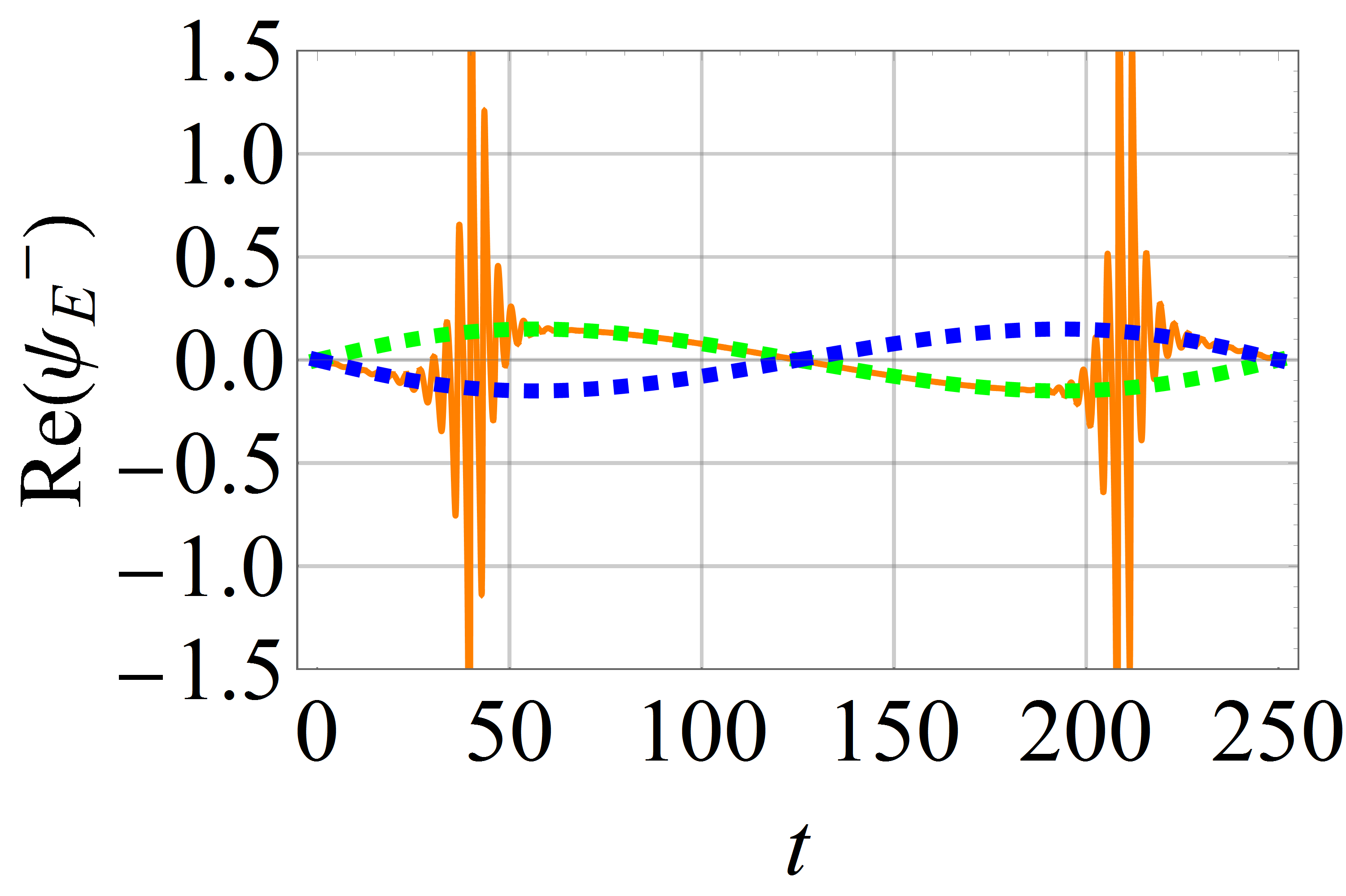}&
		\includegraphics[width=0.48\columnwidth]{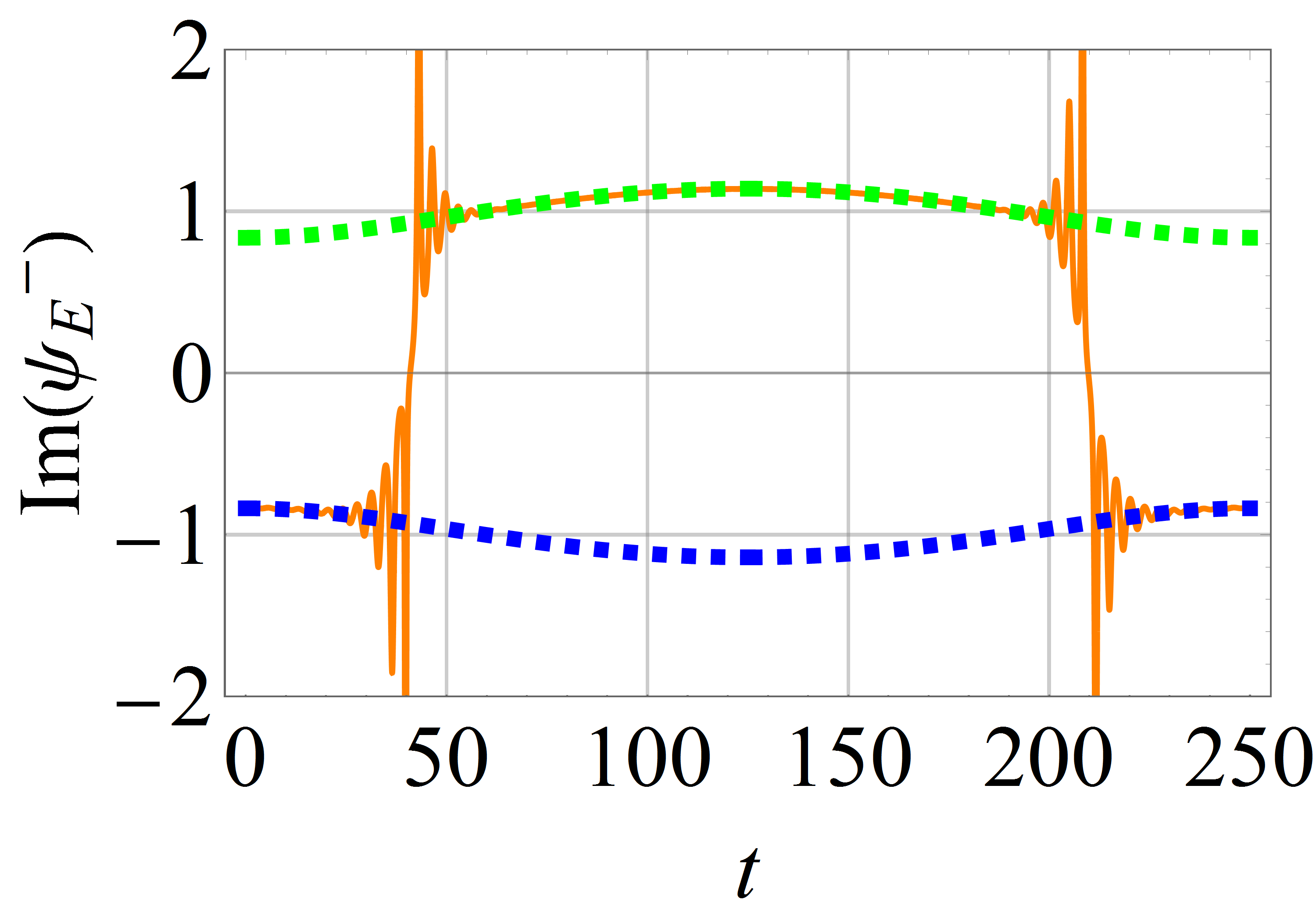}
		\end{array}$
		\caption{\label{fig:Eminus}(color online)
			Comparison between two non-cyclic states started from the energy eigenstates of the model depicted by Eq.~(\ref{eqn:BU}) for $\rho/r=0.3 $ and the two instantaneous eigenstates of $H_{\rm BU}(t)$ for sufficiently slow driving $(T=250)$. For a state $[a(t), b(t)]^T$, the plotted vertical coordinates represent the ratios [denoted $\psi_E=b(t)/a(t)$] of the two components of time-evolving non-cyclic states $U(t)|E_\pm(0)\>$ (solid lines), as compared with the parallel behavior of two instantaneous eigenstates of $H_{\rm BU}(t)$ (dotted lines). Panels (a) and (b) are for one evolving state starting from one energy eigenstate, and panels (c) and (d) are for that from the other energy eigenstate.}
	\end{center}
\label{buhop}
\end{figure}

\begin{figure}[h!]
	\begin{center}
		$\begin{array}{cc}
		\qquad\quad~~~{\rm (a)}& \qquad\qquad{\rm (b)}\\
		\includegraphics[width=0.48\columnwidth]{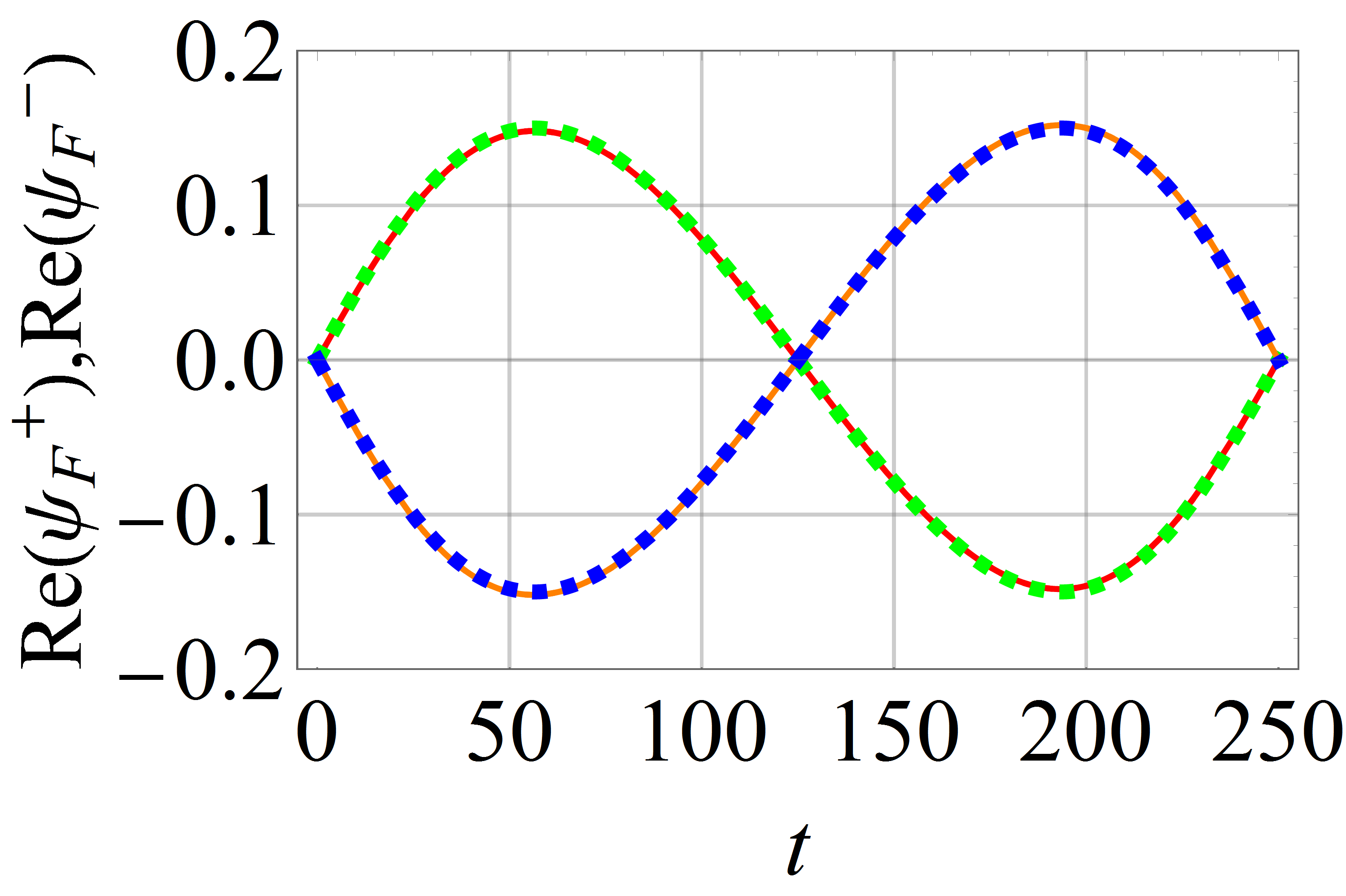}&
		\includegraphics[width=0.48\columnwidth]{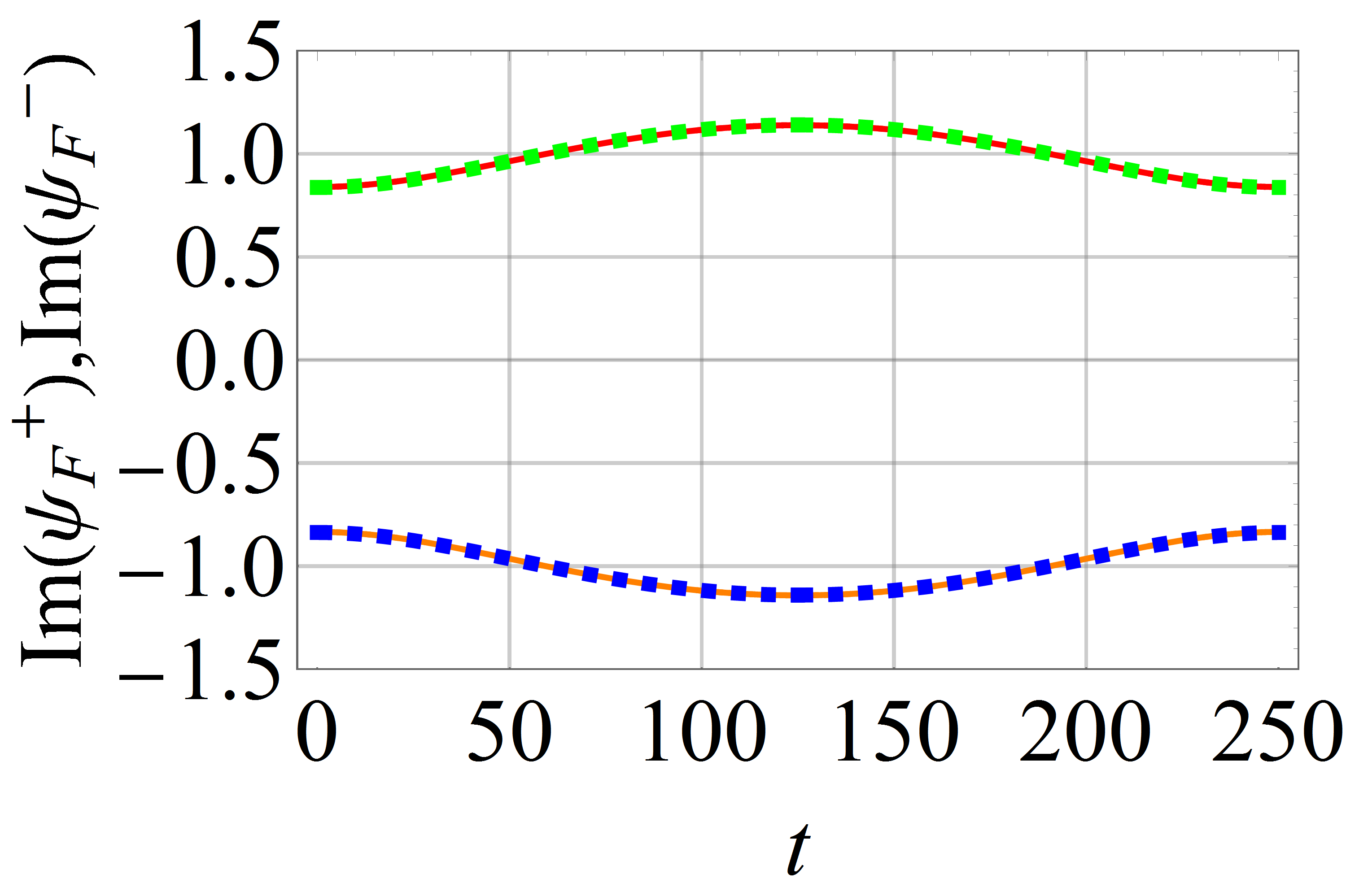}
		\end{array}$
		\caption{\label{fig:BUF}(color online) Comparison between the cyclic states of the model Eq.~(\ref{eqn:BU}) for $\rho/r=0.3$ and the instantaneous eigenstates of $H_{\rm BU}(t)$ for sufficiently slow driving $(T=250)$. For a state $[a(t), b(t)]^T$, the plotted vertical coordinates represent the ratios [denoted $\psi_F=b(t)/a(t)$] of the two components of the time-evolving cyclic states $|F^\pm(t)\>$ (solid lines), compared with two respective instantaneous eigenstates of $H_{\rm BU}(t)$ (dotted line).  Panel (a) is for the real parts of two different cyclic states, and panel (b) is for the imaginary parts. That the solid lines almost perfectly overlap with dotted lines indicates adiabatic following.}
	\end{center}
\end{figure}

In the previous subsection we have identified the piecewise adiabatic following as the consequence of a phase transition in the parameter space. For $\rho/r<(\rho/r)_{\rm crit}$, the cyclic states follow the instantaneous eigenstates in the slow driving limit. For $\rho/r>(\rho/r)_{\rm crit}$, one of the cyclic states hops (totally twice over one period) between the two instantaneous  eigenstates. The critical ratio $(\rho/r)_{\rm crit}$ is the boundary of the two phases.

It is worthwhile further distinguishing our findings from previous studies. Indeed, as observed in Ref.~\cite{BU,Uzdin,Rotter}, in non-Hermitian systems,  the ensuing dynamics emanating from one of the instantaneous eigenstates may also show certain hopping behavior.  However, the observed hopping in the literature is not governed by any known phase transition. To better appreciate this difference from our results, we note that even for $\rho/r<(\rho/r)_{\rm crit}$ in the Berry-Uzdin model, the time evolving state $U(t)|E_-(0)\>$ (that is, time evolution from one eigenstate of $H_{\rm BU}$ at time zero) may also show certain hopping behavior when the driving is sufficiently slow.  This feature is depicted in Fig.~\ref{fig:Eminus}. There we see $U(t)|E_+(0)\>$ displays adiabatic following whereas the other case $U(t)|E_-(0)\>$ displays hopping, a situation well known from the literature.    By sharp contrast, as shown in Fig.~\ref{fig:BUF},  the cyclic states for $\rho/r< (\rho/r)_{\rm crit}$ can (almost perfectly) follow the instantaneous eigenstates, no matter how slow the driving is. This remarkable difference in the behavior between $U(t)|E_-(0)\>$ and $|F^-(t)\>$ further confirms what we observed in our second model. Further,  that $|F^-(t)\>$ here has adiabatic following for
$\rho/r< (\rho/r)_{\rm crit}$  also confirms what our theory predicts in the previous subsection. Remarkably,
the  hopping behavior of a non-cyclic state such as $U(t)|E_-(0)\>$ does not critically depend on the exact value of $\rho/r$.  That is, the smaller value of $\rho/r$, the longer $T$ required to cause $U(t)|E_-(0)\>$ to hop.

Consider then cases with $\rho/r>(\rho/r)_{\rm crit}$.   We examine the different hopping features between $|F^-(t)\>$ (time evolution from  a cyclic state),  $U(t)|E_-(0)\>$ (time evolution from one instantaneous eigenstate), as well as more general superposition states involving both  $|F^+(t)\>$ and $|F^-(t)\>$ (time evolution from a superposition of two cyclic states). In Table I, we compare numerically obtained hopping timing between them, using $U(t)|E_-(0)\>$,  $0.1|F^+(t)\>+|F^-(t)\>$ as well as  $0.5|F^+(t)\>+|F^-(t)\>$ as three cases with noncyclic states as the initial states. Note that the hopping takes time and the ``hopping timing'' cannot be clearly defined. In this specific model, we note that the imaginary part of the ratio of the two components of the state vanishes only once in each hopping [see the panel (d) of Fig.~\ref{fig:Eminus}]. Thus, we use it to define the hopping timing. The hopping timing for $|F^-(t)\>$ quickly converges to a fixed ratio with $T$ for sufficiently large $T$, in full agreement with our theory in the previous subsection (our theory predicts that the hopping occurs around $0.294T$). However, for the other three cases $U(t)|E_-(0)\>$, $0.1|F^+(t)\>+|F^-(t)\>$, and $0.5|F^+(t)\>+|F^-(t)\>$, their hopping timing is qualitatively different from $|F^-(t)\>$, with \textit{non-generic} characteristics.  In the slow driving limit,  the state $0.5|F^+(t)\>+|F^-(t)\>$ with similar weightage on both cyclic states tend to hop at a rather fixed absolute time, whereas both states $U(t)|E_-(0)\>$ and $0.1|F^+(t)\>+|F^-(t)\>$ display some in-between features compromising the behavior of  $|F^-(t)\>$ and  $0.5|F^+(t)\>+|F^-(t)\>$.  To understand these, we stress again that the cyclic state hops due to the Stokes phenomenon exposed above, which has a fixed limit in $\theta$ for the hopping timing (relative time). Other states such as $U(t)|E_-(0)\>$ hop due to a more obvious reason, namely, the continuous accumulation of non-adiabatic transitions (absolute time). As the driving becomes slower, the initial state  $|E_-(0)\>$ gets closer to the cyclic state, thus having a smaller projection on the ``wrong" cyclic state and hence exhibiting a different hopping timing.

To conclude, the hopping behavior of cyclic states is a true critical phenomenon, but the hopping arising from other non-cyclic states is not and hence exhibiting non-generic features. Finally, we note  that in all these cases, hopping occurs with the system parameter far away from the exception (degeneracy) point.

\onecolumngrid
\begin{center}
	\begin{table}[h!]
		\begin{center}
			\begin{tabular}{ccccc}
				\hline
				\hline
				Period & $|F^-(t)\>$ & $U(t)|E_-(0)\>$& $\left[0.1|F^+(t)\>+|F^-(t)\>\right]$  & $\left[0.5|F^+(t)\>+|F^-(t)\>\right]$ \\
				\hline
				 20 & $~\,0.376T\approx 7.5\quad$ & $~\,0.286T\approx 5.7\quad$ & $0.243T\approx 4.9~\,$ & $0.153T\approx3.1$  \\
				\hline
				 50 & $0.262T\approx13.1$ & $0.202T\approx10.1$ & $0.162T\approx 8.1~\,$ & $0.083T\approx4.2$ \\
				\hline
				100 & $0.297T\approx29.7$ & $0.150T\approx15.0$ & $0.113T\approx11.3$ & $0.051T\approx5.1$ \\
				\hline
				150 & $0.298T\approx44.7$ & $0.127T\approx19.1$ & $0.086T\approx12.9$ & $0.051T\approx7.6$ \\
				\hline
				200 & $0.299T\approx59.7$ & $0.115T\approx23.0$ & $0.071T\approx14.1$ & $0.042T\approx8.3$ \\
				\hline
				250 & $0.302T\approx75.4$ & $0.106T\approx26.5$ & $0.068T\approx16.9$ & $0.035T\approx8.8$ \\
				\hline
				\hline
			\end{tabular}
		\end{center}
		\caption{\label{tab:hopping}Hopping timing for three types of states. Here the hopping timing is defined subjectively as when the imaginary part of $b/a$ vanishes, where the state vector is $(a,b)^T$. The parameters used are $r=1$ and $\rho=0.5$.}

	\end{table}
\end{center}
\twocolumngrid

\section{Conclusions}
\label{sec:conclusions}

In this work, we have analyzed the adiabatic following dynamics of periodically driven non-Hermitian systems. We have uncovered a new possibility of adiabatic following when considering the time evolution of cyclic states in terms of instantaneous eigenstates of the system.  As seen from our second and third models, the adiabatic following can be \textit{piecewise} due to a hopping behavior in representation of instantaneous eigenstates. We have shown that the hopping behavior can occur with system parameters far from the degeneracy (exceptional) point. In the third (exactly solvable) model, the piecewise adiabatic following therein is found to have an underlying phase boundary in the parameter space. It is hence a genuine critical behavior.  The phase boundary is also located by a straightforward asymptotic analysis.

As a side contribution, we have also advocated to use the AA phase to characterize and understand the geometrical aspects of adiabatic following in non-unitary dynamics.  Without using any sophisticated terminologies, we have shown that the AA phase we propose to use and likewise the Berry phase in the case of perfect adiabatic following are always real in non-unitary dynamics. In particular, we have shown that an earlier expression for AA phase in Hermitian systems can also apply to non-unitary dynamics (with normalized initial states). From our results, it becomes clear now that previous studies suggesting complex AA phases, though interesting in their own right, are not really consistent with the simple notion that the AA phase for non-unitary dynamics should just reflect the geometry of a closed curve in a projective Hilbert space.  As detailed in two models, if adiabatic following with instantaneous eigenstates persists without hopping, then the AA phase expectedly reduces to the Berry phase in the slow driving limit. However, for piecewise adiabatic following,  the AA phase behavior is found to be extremely complicated, suggesting rich geometrical features of non-unitary dynamics.

\textit{Additional note}: After we first reported the hopping behavior in our preprint \cite{wgold},  a second study  reporting a hopping behavior also away from exceptional points was published \cite{prepa}. However, the hopping behavior in Ref.~\cite{prepa} is not about cyclic states and hence does not have the critical features as exposed here.

\vspace{0.5cm}
\section*{Acknowledgments}
Q.W.~thanks Prof.~Yogesh N.~Joglekar and Prof.~Gerardo Ortiz for the interesting discussions and suggestions. Q.W.~also thanks Prof.~Ali Mostafazadeh for comments and Ref.~\cite{Ali99}. J.G.~is supported by Singapore Ministry of Education Academic Research Fund Tier I (WBS No.~R-144-000-353-112) and by the Singapore NRF grant No. NRF-NRFI2017-04 (WBS No. R-144-000- 378-281).   Q.W.~is supported by Singapore Ministry of Education Academic Research Fund Tier I (WBS No.~R-144-000-352-112).


\end{document}